\documentclass[%
 reprint,
superscriptaddress,
reprint,
 amsmath,amssymb,
 aps,
prl,
floatfix,
]{revtex4-2}

\usepackage{booktabs}
\usepackage{graphicx}
\usepackage{dcolumn}
\usepackage{bm}
\usepackage{hyperref}
\usepackage[dvipsnames]{xcolor}
\usepackage{soul}
\usepackage{ulem}
\usepackage{feynmp-auto}
\usepackage{mathtools}
\usepackage{physics}

\DeclareMathAlphabet{\mathpzc}{OT1}{pzc}{m}{it}

\hypersetup{
    colorlinks=true,
    linkcolor=blue,
    filecolor=blue,      
    urlcolor=blue,
    citecolor=blue
    }

\DeclareMathAlphabet{\mathpzc}{OT1}{pzc}{m}{it}

\usepackage{lipsum}
\usepackage[dvipsnames]{xcolor}
\usepackage{soul}
\newcommand{\p}{\partial}

\usepackage{mathtools}
\usepackage{color}
\usepackage{ulem}

\newcommand{\cD} {\mathcal{D}}

\begin{document}

\title{Orbital Inverse Faraday and Cotton-Mouton Effects in Hall Fluids}

\author{Gabriel Cardoso}
\affiliation{Nordita, Stockholm University, and KTH Royal Institute of Technology,
Hannes Alfv\'ens v\"ag 12, SE-106 91 Stockholm, Sweden}

\author{Erlend Syljuåsen}
\affiliation{Nordita, Stockholm University, and KTH Royal Institute of Technology,
Hannes Alfv\'ens v\"ag 12, SE-106 91 Stockholm, Sweden}
\affiliation{Center for Quantum Spintronics, Department of Physics, Norwegian University of Science and Technology, NO-7491 Trondheim, Norway}

\author{Alexander V. Balatsky}
\affiliation{Nordita, Stockholm University, and KTH Royal Institute of Technology,
Hannes Alfv\'ens v\"ag 12, SE-106 91 Stockholm, Sweden}
\affiliation{Department of Physics, University of Connecticut, Storrs, Connecticut 06269, USA}


\begin{abstract} 
    We report two light-induced orbital magnetization effects in quantum Hall (QH) fluids, stemming from their transverse response. The first is a purely transverse contribution to the inverse Faraday effect (IFE), where circularly polarized light induces a DC magnetization by stirring the charged fluid. This contribution dominates the IFE in the QH regime. The second is the orbital inverse Cotton-Mouton effect (ICME), in which linearly polarized light generates a DC magnetization. Since the applied field in the ICME does not break time-reversal symmetry, the induced magnetization directly probes the chiral orbital response of the fluid at the driving frequency. We estimate that the resulting magnetization lies in the range of $0.5$–$10$ Bohr magnetons per charge carrier in materials such as graphene and transition-metal dichalcogenides (TMDs) in the QH regime. Finally, we show that the induced magnetization is accompanied by a local correction to the static particle density, enabling optical quantum printing of density profiles into the QH fluid.
\end{abstract}

\maketitle



\noindent\textit{Introduction}-- Recent progress in control of the dynamics of quantum states has been noted.  The coupling of terahertz light to the low-energy degrees of freedom in materials enables the creation and control of quantum states on ultrafast timescales \cite{de2021colloquium}. Applications of structured light can create, modify, or destroy magnetic \cite{afanasiev2021ultrafast,mentink2017manipulating,kampfrath2011coherent}, superconducting \cite{luo2023quantum,eckhardt2024theory}, and charge-density orders \cite{kogar2020light}. One can now use the light-matter coupling to realize \textit{quantum printing}, ie., to write information into the quantum state \cite{review}. Examples are the transduction of quantum numbers into topological defects in superconductors \cite{yeh2024structured,yeh2024structured2} and magnetic materials \cite{fujita2017ultrafast}.

\begin{figure}
\centering
\includegraphics[width=1.00\linewidth]{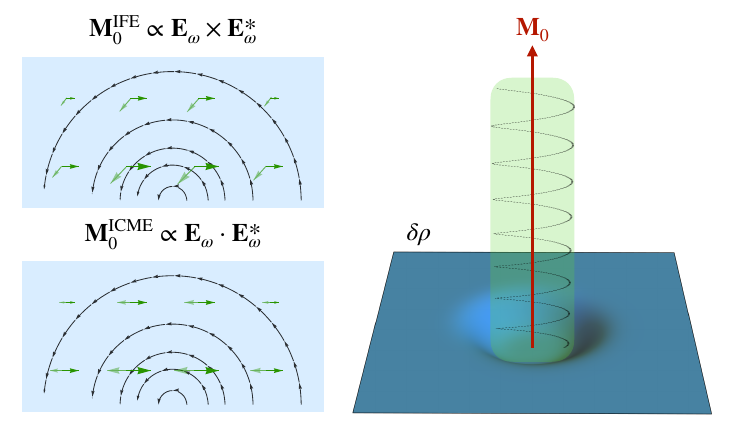}
\caption{Induction of a DC magnetization current ${\bf J}_0=\nabla\times{\bf M}_0$ (black) by finite-frequency light fields ${\bf E}_\omega$ (green) in a Hall fluid. In the inverse Faraday effect (IFE), the orbital magnetization ${\bf M}_0^{\rm IFE} \propto {\bf E}_\omega \times {\bf E}_\omega^*$ arises from stirring the fluid with circularly polarized light (top left). This effect can arise from purely transverse (Hall) response. In the inverse Cotton-Mouton effect (ICME), ${\bf M}_0^{\rm ICME} \propto {\bf E}_\omega \cdot {\bf E}_\omega^*$ is generated by linearly polarized light (bottom left), revealing the fluid’s intrinsic chiral orbital motion at the frequency of light. In the quantum Hall regime, the resulting DC magnetization is accompanied by a change of the local particle density $\delta\rho$ (right), enabling optical quantum printing of density patterns.
}
\label{fig:AIFEbeam}
\end{figure}

An important example of quantum printing is the inverse Faraday effect (IFE) \cite{pitaevskii1961electric}, where angular momentum is transferred from circularly polarized light to matter. Its signature is the generation of a net static (DC) magnetization,
\begin{equation}
    {\bf M}_0\propto{\bf E}_\omega\times{\bf E}_{\omega}^*.\label{eq:IFEdef}
\end{equation}
Here, ${\bf E}_\omega$ denotes the complex electric field of monochromatic light with frequency $\omega$. Magnetization in IFE is nonzero only for a circularly polarized light ${\bf E} \sim \hat{\bf e}_x+i\hat{\bf e}_y$, reflecting the handedness of light that is imprinted on the electron fluid. The IFE is also a prominent example of rectification, where oscillating fields induce a static response. As was realized early  \cite{pitaevskii1961electric,van1965optically}, the IFE can be generated by various microscopic mechanisms. One example is the generation of spin polarization via spin–orbit coupling \cite{taguchi2011theory,qaiumzadeh2016theory}. Alternatively, circularly polarized light can generate a magnetization by driving coherent orbital motion of charge carriers. The resulting circulating currents can generate a net static magnetization even in the absence of spin-orbit coupling (see Fig.~\ref{fig:AIFEbeam}). The orbital contribution was shown to be relevant in metals and superconductors \cite{sharma2024light,majedi2021microwave}, and the elegant description using hydrodynamics of a  charged fluid was developed \cite{hertel2006theory}. For the subsequent discussion we also recall the inverse Cotton-Mouton effect (ICME), that describes the appearance of rectified magnetization upon incidence of linearly polarized light, ${\bf M}_0\propto {\bf E}_\omega\cdot{\bf E}_\omega^*$ \cite{pershan1966theoretical,kirilyuk2010ultrafast}. 

In this letter, we extend the hydrodynamics  of electron fluids interacting with light to  develop orbital magnetization effects in charged fluids with a transverse response. In particular, we find: (i) a purely transverse contribution to the IFE. We find that the transverse (Hall) response of the fluid allows for IFE and the result depends on the Hall conductivity. Together with the longitudinal contribution, the net magnetization is:
\begin{equation}
    {\bf M}_0 = i\frac{|\sigma_\omega^L|^2+|\sigma_\omega^H|^2}{e\omega\rho_0}{\bf E}_\omega\times{\bf E}_\omega^*,\label{eq:IFEcoeff}
\end{equation}
where $e$ is the charge of carriers in the fluid, $\rho_0$ is the particle number density, and $\sigma_\omega^{L(H)}$ is the longitudinal (Hall) conductivity, respectively.
(ii)  We also predict an orbital inverse Cotton-Mouton effect (ICME), the generation of a static magnetization by linearly polarized light. In the orbital ICME the induced magnetization is given by
\begin{equation}
    {\bf M}_0=\frac{i}{e\omega\rho_0}{\rm Tr}(\sigma_\omega^{\mathrm{2D}}\times\sigma_\omega^{\mathrm{2D}*}) {\bf E}_\omega\cdot {\bf E}_\omega^*\,{\bf\hat{k}},\label{eq:AIFEdot}
\end{equation}
where $\bf\hat k$ is the direction of light propagation, $\sigma_\omega^{\mathrm{2D}}$ denotes the optical conductivity tensor on the plane perpendicular to ${\bf\hat{k}}$, and $(\sigma_\omega^{\mathrm{2D}}\times\sigma_\omega^{\mathrm{2D}*})_{ij}=\varepsilon^{kl}\sigma_{\omega,ki}^{\mathrm{2D}}\sigma_{\omega,lj}^{\mathrm{2D}*}$. This coefficient is only non-vanishing if the orbital motion of charges in the material breaks time-reversal symmetry (TRS), in contrast with the IFE \eqref{eq:IFEdef}, where TRS is broken by the circular polarization of light itself. In isotropic fluids it translates to a phase difference between the longitudinal and transverse AC responses (see Fig.~\ref{fig:AIFEbeam}). (iii) We estimate the strength of these effects in the quantum Hall (QH) fluid. In graphene and transition-metal dichalcogenides (TMDs), they lead to a significant local magnetization in the range of $0.5-10$ Bohr magnetons per charge carrier, which can be experimentally detected via the light-induced anomalous Hall effect. (iv)  We also find that in the QH fluid the induced magnetization is accompanied by a local change of the particle density. This is a modification of the Středa formula, which relates the density to the magnetic flux, by a back-reaction of the induced magnetization. As a result, light can be used to realize quantum Hall printing by optically writing density patterns into the fluid, as illustrated in Fig.~\ref{fig:AIFEbeam}.
We now proceed to elaborate these claims.


\vspace{.5cm}
    \noindent\textit{Hydrodynamics}--Our approach is based on the hydrodynamics of a charged fluid, similar to the original analysis by Pitaevskii \cite{pitaevskii1961electric} of the drift currents in plasmas. This approach is valid in the high-frequency optical regime, where the charge carriers can be treated as collisionless on the timescale of the optical field. We assume that the electric current transported by the fluid is given by the constitutive relation
\begin{equation}
    {\bf J} = e\rho{\bf v},\label{eq:Jrhov}
\end{equation}
where $\rho(t,{\bf x})$ and ${\bf v}(t,{\bf x})$ are the density and velocity fields of the fluid. Under an applied monochromatic high-frequency electromagnetic wave with electric field ${\bf E}(\bm{r}, t) = {\bf E}_0(\bm{r}) \cos(\omega t)$, the time-dependence of the fluid properties can be expanded in terms of their harmonic components,
\begin{equation}
    \rho(t,{\bf x})=\rho_0+\rho_\omega e^{i\omega t}+\rho_{-\omega} e^{-i\omega t}+\rho_{2\omega} e^{i2\omega t}+...,
\end{equation}
and similarly for $\bf v$ and $\bf J$. The reality condition of these quantities implies that $\rho_{n\omega}^{*}=\rho_{-n\omega}$. It then follows from the constitutive relation \eqref{eq:Jrhov} that
\begin{subequations}
\begin{align}
    {\bf J}_0 &= e(\rho_{-\omega} {\bf v}_\omega + \rho_\omega {\bf v}_{-\omega}) + \dots, \label{eq:J0}
    \\
    {\bf J}_\omega &= e \rho_0 {\bf v}_\omega + \dots, \label{eq:J1}
\end{align}
\end{subequations}
where we assumed that the fluid is at rest on average, ${\bf v}_0=0$. The fluctuations of the density are fixed through the continuity equation
\begin{equation}
    \p_t(e\rho)+\nabla\cdot{\bf J}=0, \label{eq:continuity}
\end{equation}
from which we obtain
\begin{align}
    \rho_{\omega}=\frac{i}{e\omega}{\bf\nabla\cdot J}_\omega.\label{eq:rho1}
\end{align}
Substituting Eqs. (\ref{eq:J1}, \ref{eq:rho1}) into Eq. \eqref{eq:J0} yields the leading order contribution to the static current,
\begin{align}
	{\bf J}_0 = \frac{i}{e \omega \rho_0} \left[({\bf\nabla\cdot J_\omega}){\bf J}_{-\omega}-({\bf\nabla\cdot J_{-\omega}}){\bf J}_{\omega}\right],
\end{align}
which can be expressed as
\begin{equation}
    {\bf J}_0 =\nabla\times {\bf M}_0+ {\bf\Gamma}_0.
\end{equation}
The static magnetization
\begin{equation}
	{\bf M}_0 = \frac{i}{e \omega \rho_0}{\bf J}_\omega\times {\bf J}_{\omega}^*\label{eq:MJJ}
\end{equation}
arises by rectification of circulating currents ${\bf J}_\omega$ present in the lower-order response. The additional contribution
\begin{equation}
    {\bf\Gamma}_0={\rm Re}\left[\frac{i}{e \omega \rho_0}({\bf J}_{-\omega}\cdot\nabla){\bf J}_\omega\right],
\end{equation}
leads to a net shift ${\bf v} \to {\bf v} + {\bf \Gamma}_0 / (e \rho_0)$ of the fluid velocity, that determines the ponderomotive forces on the fluid \cite{karpman1982ponderomotive}.

\vspace{.5cm}
\noindent\textit{Symmetry analysis}--We now consider the particular case of magnetization generated along the direction of light propagation. Then Eq.~\eqref{eq:MJJ} is expressed as
\begin{equation}
    M_0 = \frac{i}{e\omega\rho_0}(\sigma_\omega^{\rm 2D}\times\sigma_\omega^{\rm{2D}*})_{kl}E_\omega^kE_{\omega}^{l*},
\end{equation}
where only the components of the conductivity tensor on the plane perpendicular to the propagation direction appear. Recall that a rank-two tensor can be decomposed into a scalar part, an antisymmetric part, and a quadrupolar (traceless symmetric) part. Applying this decomposition to the $\sigma\times\sigma^*$ matrix, we find that its fully antisymmetric part couples to $\bf E\times E^*$, resulting in the generalized expression for the IFE,
\begin{equation}
    {\bf M}_0 = \frac{i}{2e\omega\rho_0}\varepsilon^{ij}(\sigma_\omega^{2D}\times\sigma_\omega^{2D*})_{ij}{\bf E}_\omega\times{\bf E}_\omega^*.\label{eq:IFEcross}
\end{equation}
For an isotropic material, $(\sigma_\omega^{2D})_{ij}=\sigma^L_\omega\delta_{ij}+\sigma^H_\omega\varepsilon_{ij}$, it simplifies to Eq.(\ref{eq:IFEcoeff}).

In the $\sigma^H_\omega\to 0$ case, equation \eqref{eq:IFEcoeff} reduces to the usual formula for the IFE in a longitudinal fluid \cite{hertel2006theory}. Conversely, in quantum Hall fluids it is the $\sigma^H_\omega$ contribution which dominates (see table \ref{tab:effects}). Note that, although the Hall conductivity is odd under time-reversal, its absolute value squared, appearing in the coefficient of \eqref{eq:IFEcoeff}, is even.

Consider now the symmetric part of $\sigma_\omega^{2D} \times \sigma_\omega^{2D*}$. Taking the trace yields the ICME contribution \eqref{eq:AIFEdot}. The ICME is most commonly studied in magnetic materials, where spin polarization is induced by light \cite{pershan1966theoretical,kirilyuk2010ultrafast}. This is because linearly polarized light—unlike circularly polarized light—is invariant under TRS, and therefore cannot generate magnetization unless the material response itself breaks TRS. In our case, however, the fluid is not magnetically ordered and does not necessarily break TRS at the DC level. Instead, it exhibits chiral orbital motion at the frequency of light. This orbital chirality is encoded in the specific combination of components of the conductivity tensor appearing in the coefficient of \eqref{eq:AIFEdot}. In the isotropic case, it simplifies to
\begin{equation}
    M_0=\frac{2{\rm Im}(\sigma_\omega^L\sigma_\omega^{H*})}{e\omega\rho_0}{\bf E}_\omega\cdot{\bf E}_\omega^*.\label{eq:tracecoeff}
\end{equation}
This shows that both longitudinal and transverse responses are required, along with a phase difference between them. The connection with the chiral orbital motion can be seen as follows: assume real $\sigma_\omega^H$ and imaginary $\sigma_\omega^L$ (as is the case in the QH plateaus), and let ${\bf E} = E \cos(\omega t)\hat{\bf e}_x$. Then linear response gives the current
\begin{align}
    {\bf J} = E(i\sigma^L_\omega\sin(\omega t)\hat{\bf e}_x-\sigma^H_\omega\cos(\omega t)\hat{\bf e}_y),
\end{align}
which rotates with chirality determined by the sign of ${\rm Im}(\sigma^L_\omega\sigma^{H*}_\omega)$. According to equation \eqref{eq:tracecoeff}, this also determines the sign of the static magnetization induced at the level of non-linear response (compare with Fig.~\ref{fig:AIFEbeam}). We summarize the contrast between the symmetry properties of the IFE and the ICME in table \ref{tab:symmetry}.

\begin{table}[]
        \begin{tabular}{cccc}
            \toprule
            field component\,\,\, & fluid response coefficients\,\,\, & effect\,\, \\
            \midrule
            ${\bf E}_\omega\times {\bf E}_\omega^*$ (TRS-odd) & $|\sigma_\omega^L|^2+|\sigma_\omega^H|^2$ (TRS-even) & IFE \\
            ${\bf E}_\omega\cdot {\bf E}_\omega^*$ (TRS-even) & Im$(\sigma_\omega^L\sigma_\omega^{H*})$ (TRS-odd) & ICME \\
            \bottomrule
        \end{tabular}
        \caption{Rectified magnetization effects in an isotropic charged fluid. While in the IFE the applied field breaks TRS, in the ICME it is the chiral orbital response which breaks TRS. Both effects are observable in quantum Hall systems (see table \ref{tab:effects}).}
         \label{tab:symmetry}
\end{table}

\vspace{.5cm}
\noindent\textit{Quantum Hall fluids}--The quantum Hall (QH) system is a prime example of an isotropic fluid with large transverse response, and we find that it realizes both the transverse IFE and the ICME. The hydrodynamic description of QH states can be derived from the action principle \cite{stone1990superfluid,abanov2013effective}
\begin{equation}
    S = -\int d^2x\left[\rho\left(\p_t\theta+eA_t+\frac{m{\bf v}^2}{2}\right)+V(\rho)\right],\label{eq:QHaction}
\end{equation}
where $e$ and $m$ are the particles' charge and mass, respectively, and we assume the Landau level filling fraction $\nu$ to correspond to one of the integer or fractional QH plateaus. There are two fluctuating fields: the particle density $\rho$ and the velocity potential $\theta$, which parametrizes the local velocity of the fluid ${\bf v}(t,{\bf x})$ as
\begin{equation}
    m {\bf v}=\nabla \theta+\frac{e}{c}{\bf A}+\frac{2\pi\hbar}{\nu}\nabla^\perp\left(\frac{\rho}{\nabla^2}\right).\label{eq:vHall}
\end{equation}
Here we introduce convenient notations: $v^\perp_i=\varepsilon_{ij}v^j$ denotes the rotation of vector $\bf v$ by $\pi/2$, and $\frac{e\rho}{\nabla^2}({\bf x})=\int d^2x'\frac{\log|{\bf x}-{\bf x}'|}{2\pi}e\rho({\bf x}')$ denotes the 2D electrostatic potential generated by the charge distribution $e\rho$. This action can be derived from the Chern-Simons-Landau-Ginzburg theory of quantum Hall effect \cite{zhang1992chern}, and captures the universal transport features of both integer and fractional quantum Hall states at large scales.

Assuming a constant background magnetic field $B_0$, the quantum Hall state corresponds to the homogeneous, static solution of the equations of motion. One finds that such a solution exists only for the specific value of density given by the Středa relation $\rho_0=\nu\frac{eB_0}{2\pi\hbar}$. As we review in the supplementary material \cite{supplementary} (see also \cite{abanov2013effective}), adding a perturbation to the background field $\delta A_\mu$ and expanding the effective action to linear order around the static solution, one recovers the formulas for the AC conductivities at the center of the Hall plateau,
\begin{align}
    &\sigma^H_\omega=\nu\frac{e^2}{h}\frac{\omega_0^2}{\omega_0^2-\omega^2}, &\sigma^L_\omega=\nu\frac{e^2}{h}\frac{i\omega\omega_0}{\omega^2-\omega_0^2},
\end{align}
where $\omega_0 = \frac{eB_0}{m}$ is the cyclotron frequency. Importantly, the hydrodynamic action \eqref{eq:QHaction} should be seen as the leading part of a small-gradient/low-frequency expansion with respect to $\omega_0$. We then find that the leading contribution to the IFE arises from the transverse component $\propto|\sigma_\omega^H|^2$ and is given by
\begin{equation}
    {\bf M}_0= \frac{i\nu^2e^3}{h^2\omega\rho_0}{\bf E}_\omega\times{\bf E}_\omega^*.\label{eq:IFEhydro}
\end{equation}
Similarly, the leading contribution to the ICME is
\begin{equation}
    {\bf M}_0= -\frac{\nu^2e^3}{h^2\omega_0\rho_0}{\bf E}_\omega\cdot{\bf E}_\omega^*\,\,\hat{\bf k}.\label{eq:AIFEhydro}
\end{equation}
Note that in this regime the IFE is larger than the ICME by a factor of $\omega_0/\omega$ (see Table \ref{tab:effects}). These formulas are confirmed by directly computing the non-linear response from the effective action
\begin{equation}
    S_{\rm eff}[A] = -i\log \int \cD\rho\cD\theta e^{iS[\rho,\theta;A]},
\end{equation}
as we show in the supplementary material \cite{supplementary}. We also investigate the opposite regime $\frac{\omega}{\omega_0}\gg 1$, where one can derive the effective Floquet Hamiltonian of the two-dimensional electron gas (2DEG) through a high-frequency expansion. Interestingly, we find that the result of this calculation is also matched by the general formulas (\ref{eq:IFEcoeff},\ref{eq:tracecoeff}), when expanded for high frequencies, up to a dimensionless prefactor \cite{supplementary}.

\vspace{.5cm}
\noindent\textit{Material estimates}--The QH effect can be realized in devices with optically accessible surfaces, allowing direct coupling of light to the 2DEG, notably in graphene \cite{hafez2018extremely,session2025optical} and TMDs \cite{jing2021terahertz}. THz light is ideal for probing the orbital IFE and ICME in the hydrodynamic regime, as it is generally below but comparable to the cyclotron frequency in these materials. Assuming a background field of $B_0=10{\rm T}$ (which corresponds to $\omega_0\approx 1.8 {\rm THz}$ for bare electrons), and applied light with frequency $\omega = 1{\rm THz}$ and $|{\bf E}_\omega|=10^5{\rm V/m}$, we find that the induced magnetization is in the range of $0.5-10$ Bohr magneton per charge carrier (see Table \ref{tab:effects}). Note that the induced orbital magnetization directly corresponds to a vortical current, which can be measured in terms of the light-induced anomalous Hall effect \cite{mciver2020light}.

\begin{table}[]
        \begin{tabular}{ccccc}
            \toprule
             & $\rho_0(10^{11}{\rm cm}^{-2})$ & $m_*/m_e$ & $M_{\rm IFE}(\mu_B\rho_0)$ & $M_{\rm ICME}(\mu_B\rho_0)$ \\
            \midrule
            ${\rm MoS}_2$ & 9.1 & $0.49$ & $0.57$ & $0.16$ \\
            Graphene & 2.0 & $0.007$ & $10.1$ & $0.04$ \\
            GaAs & 2.7 & $0.07$ & $5.5$ & $0.21$ \\
            \bottomrule
        \end{tabular}
        \caption{Material estimates for induced magnetization (in $\mu_B$ per carrier) in the quantum Hall regime at background $B=10{\rm T}$. The light field is at frequency $\omega=1{\rm THz}$ and intensity $|{\bf E}_\omega|=10^5{\rm V/m}$.}
         \label{tab:effects}
\end{table}

\vspace{.5cm}
\noindent\textit{Quantum Hall printing}--Up to this point, we only used the transport properties of the quantum Hall fluid. In other words, equations (\ref{eq:IFEcoeff},\ref{eq:tracecoeff}) apply even to classical transverse fluids. We turn now to a purely quantum feature of the QH state, namely the Středa formula for the proportionality between the particle density and the magnetic flux. A natural question is whether the magnetization generated by the light through rectification effects feeds back into the Středa formula as an additional magnetic flux. To answer this question, we consider the time-averaged (DC) part of the density, determined from the effective action by $\rho = -\frac{1}{e} \frac{\delta}{\delta A_t}S_{\rm eff}[A]$. As we show in the supplementary material \cite{supplementary}, the nonlinear corrections are related to the magnetization $M = \frac{\delta }{\delta B}S_{\rm eff}[A]$ by
\begin{equation}
    \rho = \frac{\nu e}{2\pi\hbar}B_0 - \frac{1}{e\hbar^2\omega_0}\nabla^2 M,\label{eq:Středa}
\end{equation}
where $M$ denotes the total DC magnetization normal to the Hall plane, including both the IFE and ICME contributions. From the perspective of quantum printing \cite{review}, it implies that one can use light to write into the QH states by imprinting static density textures through the IFE and the ICME (see Fig.~\ref{fig:AIFEbeam}).

Note that the magnetization contribution in \eqref{eq:Středa} depends on the sharpness of the applied field profile through the Laplacian operator, which is typically constrained by the diffraction limit. Estimating $\Delta M\approx-\frac{\hbar^2}{\lambda^2}M$, where $\lambda$ is the wavelength of the incident light, we find that the correction to the density is 
\begin{equation}
    \delta\rho\approx \left(\frac{\ell_0}{\lambda}\right)^2\frac{m_*}{m_e}\frac{M}{\mu_B},\label{eq:deltarhoestimate}
\end{equation}
where $\ell_0 = \sqrt{\frac{\hbar}{eB_0}}$ is the magnetic length in the external field $B_0$. For THz light and $B_0=10{\rm T}$, the ratio $(\ell_0/\lambda)$ is of the order of $10^{-5}$, so that induced density corrections are negligible. On the other hand, recent developments in scattering near-field optical microscopy (SNOM) have succeded in resolving length-scales comparable to the magnetic length in graphene quantum Hall samples \cite{dapolito2023infrared}. This is the perfect candidate to observe the quantum Hall printing of density profiles, as the estimated density correction \eqref{eq:deltarhoestimate} can be of the order of $0.1\rho_0$ (see Table \ref{tab:effects}). Importantly, it is not essential to generate circularly-polarized near-fields, since the ICME contribution which we introduced in this letter also generates a sizeable static magnetization.

\vspace{.5cm}
\noindent\textit{Conclusions}--We identified two distinct light-induced orbital magnetization effects in transverse (Hall) fluids. First, we showed that the IFE acquires an orbital contribution from the transverse response, which dominates in the QH regime. Second, we found that linearly polarized light can induce a DC magnetization via an orbital ICME, enabled by a phase difference between longitudinal and Hall conductivities. The resulting static magnetization reveals the chirality of the orbital AC response—even under time-reversal symmetric illumination. Third, we showed that this induced magnetization feeds back into the Středa relation, producing a static density shift. Together, these effects establish a mechanism for quantum Hall printing, in which light imprints density profiles into the electronic fluid.

We extended the hydrodynamic derivation of these effects to transverse fluids and developed an effective Floquet Hamiltonian approach in the complementary regime of high frequencies. We  estimated the strength of these effects in quantum Hall states in  graphene and TMDs and found that these effects can generate local orbital magnetizations of $0.5$–$10$ Bohr magnetons per carrier. Our results point to new avenues for optical control and quantum printing in topological phases, and call for additional experimental exploration of near-field light–matter coupling in QH states, where our predicted density response is enhanced by the large field gradients. Finally, this last point motivates additional theoretical studies beyond the hydrodynamic gradient expansion.

\vspace{.5cm}
\noindent\textit{Acknowledgements}--We are grateful to A. J. Millis, D. H. Lee, A. G. Abanov, and T. H. Hansson for useful discussions. This work was supported by the U.S. Department of Energy, Office of
Science, Office of Basic Energy Sciences under Award No.DE-SC-0025580 (A.V.B.), the Norwegian Research Council through Grant No. 262633, “Center of Excellence on Quantum Spintronics” (E.S.), and the European Research Council under the European Union Seventh Framework ERS-2018-SYG 810451 HERO (G.C.).

\bibliography{apssamp}

\begin{thebibliography}{32}%
\makeatletter
\providecommand \@ifxundefined [1]{%
 \@ifx{#1\undefined}
}%
\providecommand \@ifnum [1]{%
 \ifnum #1\expandafter \@firstoftwo
 \else \expandafter \@secondoftwo
 \fi
}%
\providecommand \@ifx [1]{%
 \ifx #1\expandafter \@firstoftwo
 \else \expandafter \@secondoftwo
 \fi
}%
\providecommand \natexlab [1]{#1}%
\providecommand \enquote  [1]{``#1''}%
\providecommand \bibnamefont  [1]{#1}%
\providecommand \bibfnamefont [1]{#1}%
\providecommand \citenamefont [1]{#1}%
\providecommand \href@noop [0]{\@secondoftwo}%
\providecommand \href [0]{\begingroup \@sanitize@url \@href}%
\providecommand \@href[1]{\@@startlink{#1}\@@href}%
\providecommand \@@href[1]{\endgroup#1\@@endlink}%
\providecommand \@sanitize@url [0]{\catcode `\\12\catcode `\$12\catcode `\&12\catcode `\#12\catcode `\^12\catcode `\_12\catcode `\%12\relax}%
\providecommand \@@startlink[1]{}%
\providecommand \@@endlink[0]{}%
\providecommand \url  [0]{\begingroup\@sanitize@url \@url }%
\providecommand \@url [1]{\endgroup\@href {#1}{\urlprefix }}%
\providecommand \urlprefix  [0]{URL }%
\providecommand \Eprint [0]{\href }%
\providecommand \doibase [0]{https://doi.org/}%
\providecommand \selectlanguage [0]{\@gobble}%
\providecommand \bibinfo  [0]{\@secondoftwo}%
\providecommand \bibfield  [0]{\@secondoftwo}%
\providecommand \translation [1]{[#1]}%
\providecommand \BibitemOpen [0]{}%
\providecommand \bibitemStop [0]{}%
\providecommand \bibitemNoStop [0]{.\EOS\space}%
\providecommand \EOS [0]{\spacefactor3000\relax}%
\providecommand \BibitemShut  [1]{\csname bibitem#1\endcsname}%
\let\auto@bib@innerbib\@empty
\bibitem [{\citenamefont {De~La~Torre}\ \emph {et~al.}(2021)\citenamefont {De~La~Torre}, \citenamefont {Kennes}, \citenamefont {Claassen}, \citenamefont {Gerber}, \citenamefont {McIver},\ and\ \citenamefont {Sentef}}]{de2021colloquium}%
  \BibitemOpen
  \bibfield  {author} {\bibinfo {author} {\bibfnamefont {A.}~\bibnamefont {De~La~Torre}}, \bibinfo {author} {\bibfnamefont {D.~M.}\ \bibnamefont {Kennes}}, \bibinfo {author} {\bibfnamefont {M.}~\bibnamefont {Claassen}}, \bibinfo {author} {\bibfnamefont {S.}~\bibnamefont {Gerber}}, \bibinfo {author} {\bibfnamefont {J.~W.}\ \bibnamefont {McIver}},\ and\ \bibinfo {author} {\bibfnamefont {M.~A.}\ \bibnamefont {Sentef}},\ }\bibfield  {title} {\bibinfo {title} {Colloquium: Nonthermal pathways to ultrafast control in quantum materials},\ }\href@noop {} {\bibfield  {journal} {\bibinfo  {journal} {Reviews of Modern Physics}\ }\textbf {\bibinfo {volume} {93}},\ \bibinfo {pages} {041002} (\bibinfo {year} {2021})}\BibitemShut {NoStop}%
\bibitem [{\citenamefont {Afanasiev}\ \emph {et~al.}(2021)\citenamefont {Afanasiev}, \citenamefont {Hortensius}, \citenamefont {Ivanov}, \citenamefont {Sasani}, \citenamefont {Bousquet}, \citenamefont {Blanter}, \citenamefont {Mikhaylovskiy}, \citenamefont {Kimel},\ and\ \citenamefont {Caviglia}}]{afanasiev2021ultrafast}%
  \BibitemOpen
  \bibfield  {author} {\bibinfo {author} {\bibfnamefont {D.}~\bibnamefont {Afanasiev}}, \bibinfo {author} {\bibfnamefont {J.}~\bibnamefont {Hortensius}}, \bibinfo {author} {\bibfnamefont {B.}~\bibnamefont {Ivanov}}, \bibinfo {author} {\bibfnamefont {A.}~\bibnamefont {Sasani}}, \bibinfo {author} {\bibfnamefont {E.}~\bibnamefont {Bousquet}}, \bibinfo {author} {\bibfnamefont {Y.}~\bibnamefont {Blanter}}, \bibinfo {author} {\bibfnamefont {R.}~\bibnamefont {Mikhaylovskiy}}, \bibinfo {author} {\bibfnamefont {A.}~\bibnamefont {Kimel}},\ and\ \bibinfo {author} {\bibfnamefont {A.}~\bibnamefont {Caviglia}},\ }\bibfield  {title} {\bibinfo {title} {Ultrafast control of magnetic interactions via light-driven phonons},\ }\href@noop {} {\bibfield  {journal} {\bibinfo  {journal} {Nature materials}\ }\textbf {\bibinfo {volume} {20}},\ \bibinfo {pages} {607} (\bibinfo {year} {2021})}\BibitemShut {NoStop}%
\bibitem [{\citenamefont {Mentink}(2017)}]{mentink2017manipulating}%
  \BibitemOpen
  \bibfield  {author} {\bibinfo {author} {\bibfnamefont {J.}~\bibnamefont {Mentink}},\ }\bibfield  {title} {\bibinfo {title} {Manipulating magnetism by ultrafast control of the exchange interaction},\ }\href@noop {} {\bibfield  {journal} {\bibinfo  {journal} {Journal of Physics: Condensed Matter}\ }\textbf {\bibinfo {volume} {29}},\ \bibinfo {pages} {453001} (\bibinfo {year} {2017})}\BibitemShut {NoStop}%
\bibitem [{\citenamefont {Kampfrath}\ \emph {et~al.}(2011)\citenamefont {Kampfrath}, \citenamefont {Sell}, \citenamefont {Klatt}, \citenamefont {Pashkin}, \citenamefont {M{\"a}hrlein}, \citenamefont {Dekorsy}, \citenamefont {Wolf}, \citenamefont {Fiebig}, \citenamefont {Leitenstorfer},\ and\ \citenamefont {Huber}}]{kampfrath2011coherent}%
  \BibitemOpen
  \bibfield  {author} {\bibinfo {author} {\bibfnamefont {T.}~\bibnamefont {Kampfrath}}, \bibinfo {author} {\bibfnamefont {A.}~\bibnamefont {Sell}}, \bibinfo {author} {\bibfnamefont {G.}~\bibnamefont {Klatt}}, \bibinfo {author} {\bibfnamefont {A.}~\bibnamefont {Pashkin}}, \bibinfo {author} {\bibfnamefont {S.}~\bibnamefont {M{\"a}hrlein}}, \bibinfo {author} {\bibfnamefont {T.}~\bibnamefont {Dekorsy}}, \bibinfo {author} {\bibfnamefont {M.}~\bibnamefont {Wolf}}, \bibinfo {author} {\bibfnamefont {M.}~\bibnamefont {Fiebig}}, \bibinfo {author} {\bibfnamefont {A.}~\bibnamefont {Leitenstorfer}},\ and\ \bibinfo {author} {\bibfnamefont {R.}~\bibnamefont {Huber}},\ }\bibfield  {title} {\bibinfo {title} {Coherent terahertz control of antiferromagnetic spin waves},\ }\href@noop {} {\bibfield  {journal} {\bibinfo  {journal} {Nature Photonics}\ }\textbf {\bibinfo {volume} {5}},\ \bibinfo {pages} {31} (\bibinfo {year} {2011})}\BibitemShut {NoStop}%
\bibitem [{\citenamefont {Luo}\ \emph {et~al.}(2023)\citenamefont {Luo}, \citenamefont {Mootz}, \citenamefont {Kang}, \citenamefont {Huang}, \citenamefont {Eom}, \citenamefont {Lee}, \citenamefont {Vaswani}, \citenamefont {Collantes}, \citenamefont {Hellstrom}, \citenamefont {Perakis} \emph {et~al.}}]{luo2023quantum}%
  \BibitemOpen
  \bibfield  {author} {\bibinfo {author} {\bibfnamefont {L.}~\bibnamefont {Luo}}, \bibinfo {author} {\bibfnamefont {M.}~\bibnamefont {Mootz}}, \bibinfo {author} {\bibfnamefont {J.-H.}\ \bibnamefont {Kang}}, \bibinfo {author} {\bibfnamefont {C.}~\bibnamefont {Huang}}, \bibinfo {author} {\bibfnamefont {K.}~\bibnamefont {Eom}}, \bibinfo {author} {\bibfnamefont {J.}~\bibnamefont {Lee}}, \bibinfo {author} {\bibfnamefont {C.}~\bibnamefont {Vaswani}}, \bibinfo {author} {\bibfnamefont {Y.}~\bibnamefont {Collantes}}, \bibinfo {author} {\bibfnamefont {E.}~\bibnamefont {Hellstrom}}, \bibinfo {author} {\bibfnamefont {I.~E.}\ \bibnamefont {Perakis}}, \emph {et~al.},\ }\bibfield  {title} {\bibinfo {title} {Quantum coherence tomography of light-controlled superconductivity},\ }\href@noop {} {\bibfield  {journal} {\bibinfo  {journal} {Nature Physics}\ }\textbf {\bibinfo {volume} {19}},\ \bibinfo {pages} {201} (\bibinfo {year} {2023})}\BibitemShut {NoStop}%
\bibitem [{\citenamefont {Eckhardt}\ \emph {et~al.}(2024)\citenamefont {Eckhardt}, \citenamefont {Chattopadhyay}, \citenamefont {Kennes}, \citenamefont {Demler}, \citenamefont {Sentef},\ and\ \citenamefont {Michael}}]{eckhardt2024theory}%
  \BibitemOpen
  \bibfield  {author} {\bibinfo {author} {\bibfnamefont {C.~J.}\ \bibnamefont {Eckhardt}}, \bibinfo {author} {\bibfnamefont {S.}~\bibnamefont {Chattopadhyay}}, \bibinfo {author} {\bibfnamefont {D.~M.}\ \bibnamefont {Kennes}}, \bibinfo {author} {\bibfnamefont {E.~A.}\ \bibnamefont {Demler}}, \bibinfo {author} {\bibfnamefont {M.~A.}\ \bibnamefont {Sentef}},\ and\ \bibinfo {author} {\bibfnamefont {M.~H.}\ \bibnamefont {Michael}},\ }\bibfield  {title} {\bibinfo {title} {Theory of resonantly enhanced photo-induced superconductivity},\ }\href@noop {} {\bibfield  {journal} {\bibinfo  {journal} {Nature Communications}\ }\textbf {\bibinfo {volume} {15}},\ \bibinfo {pages} {2300} (\bibinfo {year} {2024})}\BibitemShut {NoStop}%
\bibitem [{\citenamefont {Kogar}\ \emph {et~al.}(2020)\citenamefont {Kogar}, \citenamefont {Zong}, \citenamefont {Dolgirev}, \citenamefont {Shen}, \citenamefont {Straquadine}, \citenamefont {Bie}, \citenamefont {Wang}, \citenamefont {Rohwer}, \citenamefont {Tung}, \citenamefont {Yang} \emph {et~al.}}]{kogar2020light}%
  \BibitemOpen
  \bibfield  {author} {\bibinfo {author} {\bibfnamefont {A.}~\bibnamefont {Kogar}}, \bibinfo {author} {\bibfnamefont {A.}~\bibnamefont {Zong}}, \bibinfo {author} {\bibfnamefont {P.~E.}\ \bibnamefont {Dolgirev}}, \bibinfo {author} {\bibfnamefont {X.}~\bibnamefont {Shen}}, \bibinfo {author} {\bibfnamefont {J.}~\bibnamefont {Straquadine}}, \bibinfo {author} {\bibfnamefont {Y.-Q.}\ \bibnamefont {Bie}}, \bibinfo {author} {\bibfnamefont {X.}~\bibnamefont {Wang}}, \bibinfo {author} {\bibfnamefont {T.}~\bibnamefont {Rohwer}}, \bibinfo {author} {\bibfnamefont {I.-C.}\ \bibnamefont {Tung}}, \bibinfo {author} {\bibfnamefont {Y.}~\bibnamefont {Yang}}, \emph {et~al.},\ }\bibfield  {title} {\bibinfo {title} {Light-induced charge density wave in late3},\ }\href@noop {} {\bibfield  {journal} {\bibinfo  {journal} {Nature Physics}\ }\textbf {\bibinfo {volume} {16}},\ \bibinfo {pages} {159} (\bibinfo {year} {2020})}\BibitemShut {NoStop}%
\bibitem [{\citenamefont {{Aeppli, Gabriel et al.}}(2025)}]{review}%
  \BibitemOpen
  \bibfield  {author} {\bibinfo {author} {\bibnamefont {{Aeppli, Gabriel et al.}}},\ }\bibfield  {title} {\bibinfo {title} {Quantum printing},\ }\href@noop {} {\bibfield  {journal} {\bibinfo  {journal} {in preparation}\ } (\bibinfo {year} {2025})}\BibitemShut {NoStop}%
\bibitem [{\citenamefont {Yeh}\ \emph {et~al.}(2024{\natexlab{a}})\citenamefont {Yeh}, \citenamefont {Yerzhakov}, \citenamefont {Horn}, \citenamefont {Raghu},\ and\ \citenamefont {Balatsky}}]{yeh2024structured}%
  \BibitemOpen
  \bibfield  {author} {\bibinfo {author} {\bibfnamefont {T.-T.}\ \bibnamefont {Yeh}}, \bibinfo {author} {\bibfnamefont {H.}~\bibnamefont {Yerzhakov}}, \bibinfo {author} {\bibfnamefont {L.~B.-V.}\ \bibnamefont {Horn}}, \bibinfo {author} {\bibfnamefont {S.}~\bibnamefont {Raghu}},\ and\ \bibinfo {author} {\bibfnamefont {A.}~\bibnamefont {Balatsky}},\ }\bibfield  {title} {\bibinfo {title} {Structured light and induced vorticity in superconductors i: Linearly polarized light},\ }\href@noop {} {\bibfield  {journal} {\bibinfo  {journal} {arXiv preprint arXiv:2407.15834}\ } (\bibinfo {year} {2024}{\natexlab{a}})}\BibitemShut {NoStop}%
\bibitem [{\citenamefont {Yeh}\ \emph {et~al.}(2024{\natexlab{b}})\citenamefont {Yeh}, \citenamefont {Yerzhakov}, \citenamefont {Horn}, \citenamefont {Raghu},\ and\ \citenamefont {Balatsky}}]{yeh2024structured2}%
  \BibitemOpen
  \bibfield  {author} {\bibinfo {author} {\bibfnamefont {T.-T.}\ \bibnamefont {Yeh}}, \bibinfo {author} {\bibfnamefont {H.}~\bibnamefont {Yerzhakov}}, \bibinfo {author} {\bibfnamefont {L.~B.-V.}\ \bibnamefont {Horn}}, \bibinfo {author} {\bibfnamefont {S.}~\bibnamefont {Raghu}},\ and\ \bibinfo {author} {\bibfnamefont {A.}~\bibnamefont {Balatsky}},\ }\bibfield  {title} {\bibinfo {title} {Structured light and induced vorticity in superconductors ii: Quantum print with laguerre-gaussian beam},\ }\href@noop {} {\bibfield  {journal} {\bibinfo  {journal} {arXiv preprint arXiv:2412.00935}\ } (\bibinfo {year} {2024}{\natexlab{b}})}\BibitemShut {NoStop}%
\bibitem [{\citenamefont {Fujita}\ and\ \citenamefont {Sato}(2017)}]{fujita2017ultrafast}%
  \BibitemOpen
  \bibfield  {author} {\bibinfo {author} {\bibfnamefont {H.}~\bibnamefont {Fujita}}\ and\ \bibinfo {author} {\bibfnamefont {M.}~\bibnamefont {Sato}},\ }\bibfield  {title} {\bibinfo {title} {Ultrafast generation of skyrmionic defects with vortex beams: Printing laser profiles on magnets},\ }\href@noop {} {\bibfield  {journal} {\bibinfo  {journal} {Physical Review B}\ }\textbf {\bibinfo {volume} {95}},\ \bibinfo {pages} {054421} (\bibinfo {year} {2017})}\BibitemShut {NoStop}%
\bibitem [{\citenamefont {Pitaevskii}(1961)}]{pitaevskii1961electric}%
  \BibitemOpen
  \bibfield  {author} {\bibinfo {author} {\bibfnamefont {L.}~\bibnamefont {Pitaevskii}},\ }\bibfield  {title} {\bibinfo {title} {Electric forces in a transparent dispersive medium},\ }\href@noop {} {\bibfield  {journal} {\bibinfo  {journal} {Sov. Phys. JETP}\ }\textbf {\bibinfo {volume} {12}},\ \bibinfo {pages} {1008} (\bibinfo {year} {1961})}\BibitemShut {NoStop}%
\bibitem [{\citenamefont {Van~der Ziel}\ \emph {et~al.}(1965)\citenamefont {Van~der Ziel}, \citenamefont {Pershan},\ and\ \citenamefont {Malmstrom}}]{van1965optically}%
  \BibitemOpen
  \bibfield  {author} {\bibinfo {author} {\bibfnamefont {J.}~\bibnamefont {Van~der Ziel}}, \bibinfo {author} {\bibfnamefont {P.~S.}\ \bibnamefont {Pershan}},\ and\ \bibinfo {author} {\bibfnamefont {L.}~\bibnamefont {Malmstrom}},\ }\bibfield  {title} {\bibinfo {title} {Optically-induced magnetization resulting from the inverse faraday effect},\ }\href@noop {} {\bibfield  {journal} {\bibinfo  {journal} {Physical review letters}\ }\textbf {\bibinfo {volume} {15}},\ \bibinfo {pages} {190} (\bibinfo {year} {1965})}\BibitemShut {NoStop}%
\bibitem [{\citenamefont {Taguchi}\ and\ \citenamefont {Tatara}(2011)}]{taguchi2011theory}%
  \BibitemOpen
  \bibfield  {author} {\bibinfo {author} {\bibfnamefont {K.}~\bibnamefont {Taguchi}}\ and\ \bibinfo {author} {\bibfnamefont {G.}~\bibnamefont {Tatara}},\ }\bibfield  {title} {\bibinfo {title} {Theory of inverse faraday effect in a disordered metal in the terahertz regime},\ }\href@noop {} {\bibfield  {journal} {\bibinfo  {journal} {Physical Review B—Condensed Matter and Materials Physics}\ }\textbf {\bibinfo {volume} {84}},\ \bibinfo {pages} {174433} (\bibinfo {year} {2011})}\BibitemShut {NoStop}%
\bibitem [{\citenamefont {Qaiumzadeh}\ and\ \citenamefont {Titov}(2016)}]{qaiumzadeh2016theory}%
  \BibitemOpen
  \bibfield  {author} {\bibinfo {author} {\bibfnamefont {A.}~\bibnamefont {Qaiumzadeh}}\ and\ \bibinfo {author} {\bibfnamefont {M.}~\bibnamefont {Titov}},\ }\bibfield  {title} {\bibinfo {title} {Theory of light-induced effective magnetic field in rashba ferromagnets},\ }\href@noop {} {\bibfield  {journal} {\bibinfo  {journal} {Physical Review B}\ }\textbf {\bibinfo {volume} {94}},\ \bibinfo {pages} {014425} (\bibinfo {year} {2016})}\BibitemShut {NoStop}%
\bibitem [{\citenamefont {Sharma}\ and\ \citenamefont {Balatsky}(2024)}]{sharma2024light}%
  \BibitemOpen
  \bibfield  {author} {\bibinfo {author} {\bibfnamefont {P.}~\bibnamefont {Sharma}}\ and\ \bibinfo {author} {\bibfnamefont {A.~V.}\ \bibnamefont {Balatsky}},\ }\bibfield  {title} {\bibinfo {title} {Light-induced orbital magnetism in metals via inverse faraday effect},\ }\href@noop {} {\bibfield  {journal} {\bibinfo  {journal} {Physical Review B}\ }\textbf {\bibinfo {volume} {110}},\ \bibinfo {pages} {094302} (\bibinfo {year} {2024})}\BibitemShut {NoStop}%
\bibitem [{\citenamefont {Majedi}(2021)}]{majedi2021microwave}%
  \BibitemOpen
  \bibfield  {author} {\bibinfo {author} {\bibfnamefont {A.~H.}\ \bibnamefont {Majedi}},\ }\bibfield  {title} {\bibinfo {title} {Microwave-induced inverse faraday effect in superconductors},\ }\href@noop {} {\bibfield  {journal} {\bibinfo  {journal} {Physical Review Letters}\ }\textbf {\bibinfo {volume} {127}},\ \bibinfo {pages} {087001} (\bibinfo {year} {2021})}\BibitemShut {NoStop}%
\bibitem [{\citenamefont {Hertel}(2006)}]{hertel2006theory}%
  \BibitemOpen
  \bibfield  {author} {\bibinfo {author} {\bibfnamefont {R.}~\bibnamefont {Hertel}},\ }\bibfield  {title} {\bibinfo {title} {Theory of the inverse faraday effect in metals},\ }\href@noop {} {\bibfield  {journal} {\bibinfo  {journal} {Journal of magnetism and magnetic materials}\ }\textbf {\bibinfo {volume} {303}},\ \bibinfo {pages} {L1} (\bibinfo {year} {2006})}\BibitemShut {NoStop}%
\bibitem [{\citenamefont {Pershan}\ \emph {et~al.}(1966)\citenamefont {Pershan}, \citenamefont {Van~der Ziel},\ and\ \citenamefont {Malmstrom}}]{pershan1966theoretical}%
  \BibitemOpen
  \bibfield  {author} {\bibinfo {author} {\bibfnamefont {P.}~\bibnamefont {Pershan}}, \bibinfo {author} {\bibfnamefont {J.}~\bibnamefont {Van~der Ziel}},\ and\ \bibinfo {author} {\bibfnamefont {L.}~\bibnamefont {Malmstrom}},\ }\bibfield  {title} {\bibinfo {title} {Theoretical discussion of the inverse faraday effect, raman scattering, and related phenomena},\ }\href@noop {} {\bibfield  {journal} {\bibinfo  {journal} {Physical review}\ }\textbf {\bibinfo {volume} {143}},\ \bibinfo {pages} {574} (\bibinfo {year} {1966})}\BibitemShut {NoStop}%
\bibitem [{\citenamefont {Kirilyuk}\ \emph {et~al.}(2010)\citenamefont {Kirilyuk}, \citenamefont {Kimel},\ and\ \citenamefont {Rasing}}]{kirilyuk2010ultrafast}%
  \BibitemOpen
  \bibfield  {author} {\bibinfo {author} {\bibfnamefont {A.}~\bibnamefont {Kirilyuk}}, \bibinfo {author} {\bibfnamefont {A.~V.}\ \bibnamefont {Kimel}},\ and\ \bibinfo {author} {\bibfnamefont {T.}~\bibnamefont {Rasing}},\ }\bibfield  {title} {\bibinfo {title} {Ultrafast optical manipulation of magnetic order},\ }\href@noop {} {\bibfield  {journal} {\bibinfo  {journal} {Reviews of Modern Physics}\ }\textbf {\bibinfo {volume} {82}},\ \bibinfo {pages} {2731} (\bibinfo {year} {2010})}\BibitemShut {NoStop}%
\bibitem [{\citenamefont {Karpman}\ and\ \citenamefont {Shagalov}(1982)}]{karpman1982ponderomotive}%
  \BibitemOpen
  \bibfield  {author} {\bibinfo {author} {\bibfnamefont {V.}~\bibnamefont {Karpman}}\ and\ \bibinfo {author} {\bibfnamefont {A.}~\bibnamefont {Shagalov}},\ }\bibfield  {title} {\bibinfo {title} {The ponderomotive force of a high-frequency electromagnetic field in a cold magnetized plasma},\ }\href@noop {} {\bibfield  {journal} {\bibinfo  {journal} {Journal of Plasma Physics}\ }\textbf {\bibinfo {volume} {27}},\ \bibinfo {pages} {215} (\bibinfo {year} {1982})}\BibitemShut {NoStop}%
\bibitem [{\citenamefont {Stone}(1990)}]{stone1990superfluid}%
  \BibitemOpen
  \bibfield  {author} {\bibinfo {author} {\bibfnamefont {M.}~\bibnamefont {Stone}},\ }\bibfield  {title} {\bibinfo {title} {Superfluid dynamics of the fractional quantum hall state},\ }\href@noop {} {\bibfield  {journal} {\bibinfo  {journal} {Physical Review B}\ }\textbf {\bibinfo {volume} {42}},\ \bibinfo {pages} {212} (\bibinfo {year} {1990})}\BibitemShut {NoStop}%
\bibitem [{\citenamefont {Abanov}(2013)}]{abanov2013effective}%
  \BibitemOpen
  \bibfield  {author} {\bibinfo {author} {\bibfnamefont {A.~G.}\ \bibnamefont {Abanov}},\ }\bibfield  {title} {\bibinfo {title} {On the effective hydrodynamics of the fractional quantum hall effect},\ }\href@noop {} {\bibfield  {journal} {\bibinfo  {journal} {Journal of Physics A: Mathematical and Theoretical}\ }\textbf {\bibinfo {volume} {46}},\ \bibinfo {pages} {292001} (\bibinfo {year} {2013})}\BibitemShut {NoStop}%
\bibitem [{\citenamefont {Zhang}(1992)}]{zhang1992chern}%
  \BibitemOpen
  \bibfield  {author} {\bibinfo {author} {\bibfnamefont {S.~C.}\ \bibnamefont {Zhang}},\ }\bibfield  {title} {\bibinfo {title} {The chern--simons--landau--ginzburg theory of the fractional quantum hall effect},\ }\href@noop {} {\bibfield  {journal} {\bibinfo  {journal} {International Journal of Modern Physics B}\ }\textbf {\bibinfo {volume} {6}},\ \bibinfo {pages} {25} (\bibinfo {year} {1992})}\BibitemShut {NoStop}%
\bibitem [{sup()}]{supplementary}%
  \BibitemOpen
  \href@noop {} {\bibinfo {title} {See supplementary material}}\BibitemShut {NoStop}%
\bibitem [{\citenamefont {Hafez}\ \emph {et~al.}(2018)\citenamefont {Hafez}, \citenamefont {Kovalev}, \citenamefont {Deinert}, \citenamefont {Mics}, \citenamefont {Green}, \citenamefont {Awari}, \citenamefont {Chen}, \citenamefont {Germanskiy}, \citenamefont {Lehnert}, \citenamefont {Teichert} \emph {et~al.}}]{hafez2018extremely}%
  \BibitemOpen
  \bibfield  {author} {\bibinfo {author} {\bibfnamefont {H.~A.}\ \bibnamefont {Hafez}}, \bibinfo {author} {\bibfnamefont {S.}~\bibnamefont {Kovalev}}, \bibinfo {author} {\bibfnamefont {J.-C.}\ \bibnamefont {Deinert}}, \bibinfo {author} {\bibfnamefont {Z.}~\bibnamefont {Mics}}, \bibinfo {author} {\bibfnamefont {B.}~\bibnamefont {Green}}, \bibinfo {author} {\bibfnamefont {N.}~\bibnamefont {Awari}}, \bibinfo {author} {\bibfnamefont {M.}~\bibnamefont {Chen}}, \bibinfo {author} {\bibfnamefont {S.}~\bibnamefont {Germanskiy}}, \bibinfo {author} {\bibfnamefont {U.}~\bibnamefont {Lehnert}}, \bibinfo {author} {\bibfnamefont {J.}~\bibnamefont {Teichert}}, \emph {et~al.},\ }\bibfield  {title} {\bibinfo {title} {Extremely efficient terahertz high-harmonic generation in graphene by hot dirac fermions},\ }\href@noop {} {\bibfield  {journal} {\bibinfo  {journal} {Nature}\ }\textbf {\bibinfo {volume} {561}},\ \bibinfo {pages} {507} (\bibinfo {year} {2018})}\BibitemShut {NoStop}%
\bibitem [{\citenamefont {Session}\ \emph {et~al.}(2025)\citenamefont {Session}, \citenamefont {Jalali~Mehrabad}, \citenamefont {Paithankar}, \citenamefont {Grass}, \citenamefont {Eckhardt}, \citenamefont {Cao}, \citenamefont {Gustavo Su{\'a}rez~Forero}, \citenamefont {Li}, \citenamefont {Alam}, \citenamefont {Watanabe} \emph {et~al.}}]{session2025optical}%
  \BibitemOpen
  \bibfield  {author} {\bibinfo {author} {\bibfnamefont {D.}~\bibnamefont {Session}}, \bibinfo {author} {\bibfnamefont {M.}~\bibnamefont {Jalali~Mehrabad}}, \bibinfo {author} {\bibfnamefont {N.}~\bibnamefont {Paithankar}}, \bibinfo {author} {\bibfnamefont {T.}~\bibnamefont {Grass}}, \bibinfo {author} {\bibfnamefont {C.~J.}\ \bibnamefont {Eckhardt}}, \bibinfo {author} {\bibfnamefont {B.}~\bibnamefont {Cao}}, \bibinfo {author} {\bibfnamefont {D.}~\bibnamefont {Gustavo Su{\'a}rez~Forero}}, \bibinfo {author} {\bibfnamefont {K.}~\bibnamefont {Li}}, \bibinfo {author} {\bibfnamefont {M.~S.}\ \bibnamefont {Alam}}, \bibinfo {author} {\bibfnamefont {K.}~\bibnamefont {Watanabe}}, \emph {et~al.},\ }\bibfield  {title} {\bibinfo {title} {Optical pumping of electronic quantum hall states with vortex light},\ }\href@noop {} {\bibfield  {journal} {\bibinfo  {journal} {Nature Photonics}\ }\textbf {\bibinfo {volume} {19}},\ \bibinfo {pages} {156} (\bibinfo {year} {2025})}\BibitemShut {NoStop}%
\bibitem [{\citenamefont {Jing}\ \emph {et~al.}(2021)\citenamefont {Jing}, \citenamefont {Shao}, \citenamefont {Fei}, \citenamefont {Lo}, \citenamefont {Vitalone}, \citenamefont {Ruta}, \citenamefont {Staunton}, \citenamefont {Zheng}, \citenamefont {Mcleod}, \citenamefont {Sun} \emph {et~al.}}]{jing2021terahertz}%
  \BibitemOpen
  \bibfield  {author} {\bibinfo {author} {\bibfnamefont {R.}~\bibnamefont {Jing}}, \bibinfo {author} {\bibfnamefont {Y.}~\bibnamefont {Shao}}, \bibinfo {author} {\bibfnamefont {Z.}~\bibnamefont {Fei}}, \bibinfo {author} {\bibfnamefont {C.~F.~B.}\ \bibnamefont {Lo}}, \bibinfo {author} {\bibfnamefont {R.~A.}\ \bibnamefont {Vitalone}}, \bibinfo {author} {\bibfnamefont {F.~L.}\ \bibnamefont {Ruta}}, \bibinfo {author} {\bibfnamefont {J.}~\bibnamefont {Staunton}}, \bibinfo {author} {\bibfnamefont {W.~J.-C.}\ \bibnamefont {Zheng}}, \bibinfo {author} {\bibfnamefont {A.~S.}\ \bibnamefont {Mcleod}}, \bibinfo {author} {\bibfnamefont {Z.}~\bibnamefont {Sun}}, \emph {et~al.},\ }\bibfield  {title} {\bibinfo {title} {Terahertz response of monolayer and few-layer wte2 at the nanoscale},\ }\href@noop {} {\bibfield  {journal} {\bibinfo  {journal} {Nature communications}\ }\textbf {\bibinfo {volume} {12}},\ \bibinfo {pages} {5594} (\bibinfo {year} {2021})}\BibitemShut {NoStop}%
\bibitem [{\citenamefont {McIver}\ \emph {et~al.}(2020)\citenamefont {McIver}, \citenamefont {Schulte}, \citenamefont {Stein}, \citenamefont {Matsuyama}, \citenamefont {Jotzu}, \citenamefont {Meier},\ and\ \citenamefont {Cavalleri}}]{mciver2020light}%
  \BibitemOpen
  \bibfield  {author} {\bibinfo {author} {\bibfnamefont {J.~W.}\ \bibnamefont {McIver}}, \bibinfo {author} {\bibfnamefont {B.}~\bibnamefont {Schulte}}, \bibinfo {author} {\bibfnamefont {F.-U.}\ \bibnamefont {Stein}}, \bibinfo {author} {\bibfnamefont {T.}~\bibnamefont {Matsuyama}}, \bibinfo {author} {\bibfnamefont {G.}~\bibnamefont {Jotzu}}, \bibinfo {author} {\bibfnamefont {G.}~\bibnamefont {Meier}},\ and\ \bibinfo {author} {\bibfnamefont {A.}~\bibnamefont {Cavalleri}},\ }\bibfield  {title} {\bibinfo {title} {Light-induced anomalous hall effect in graphene},\ }\href@noop {} {\bibfield  {journal} {\bibinfo  {journal} {Nature physics}\ }\textbf {\bibinfo {volume} {16}},\ \bibinfo {pages} {38} (\bibinfo {year} {2020})}\BibitemShut {NoStop}%
\bibitem [{\citenamefont {Dapolito}\ \emph {et~al.}(2023)\citenamefont {Dapolito}, \citenamefont {Tsuneto}, \citenamefont {Zheng}, \citenamefont {Wehmeier}, \citenamefont {Xu}, \citenamefont {Chen}, \citenamefont {Sun}, \citenamefont {Du}, \citenamefont {Shao}, \citenamefont {Jing} \emph {et~al.}}]{dapolito2023infrared}%
  \BibitemOpen
  \bibfield  {author} {\bibinfo {author} {\bibfnamefont {M.}~\bibnamefont {Dapolito}}, \bibinfo {author} {\bibfnamefont {M.}~\bibnamefont {Tsuneto}}, \bibinfo {author} {\bibfnamefont {W.}~\bibnamefont {Zheng}}, \bibinfo {author} {\bibfnamefont {L.}~\bibnamefont {Wehmeier}}, \bibinfo {author} {\bibfnamefont {S.}~\bibnamefont {Xu}}, \bibinfo {author} {\bibfnamefont {X.}~\bibnamefont {Chen}}, \bibinfo {author} {\bibfnamefont {J.}~\bibnamefont {Sun}}, \bibinfo {author} {\bibfnamefont {Z.}~\bibnamefont {Du}}, \bibinfo {author} {\bibfnamefont {Y.}~\bibnamefont {Shao}}, \bibinfo {author} {\bibfnamefont {R.}~\bibnamefont {Jing}}, \emph {et~al.},\ }\bibfield  {title} {\bibinfo {title} {Infrared nano-imaging of dirac magnetoexcitons in graphene},\ }\href@noop {} {\bibfield  {journal} {\bibinfo  {journal} {Nature nanotechnology}\ }\textbf {\bibinfo {volume} {18}},\ \bibinfo {pages} {1409} (\bibinfo {year} {2023})}\BibitemShut {NoStop}%
\bibitem [{\citenamefont {Rahav}\ \emph {et~al.}(2003)\citenamefont {Rahav}, \citenamefont {Gilary},\ and\ \citenamefont {Fishman}}]{rahav2003floquet1}%
  \BibitemOpen
  \bibfield  {author} {\bibinfo {author} {\bibfnamefont {S.}~\bibnamefont {Rahav}}, \bibinfo {author} {\bibfnamefont {I.}~\bibnamefont {Gilary}},\ and\ \bibinfo {author} {\bibfnamefont {S.}~\bibnamefont {Fishman}},\ }\bibfield  {title} {\bibinfo {title} {Effective hamiltonians for periodically driven systems},\ }\href {https://doi.org/10.1103/PhysRevA.68.013820} {\bibfield  {journal} {\bibinfo  {journal} {Phys. Rev. A}\ }\textbf {\bibinfo {volume} {68}},\ \bibinfo {pages} {013820} (\bibinfo {year} {2003})}\BibitemShut {NoStop}%
\bibitem [{\citenamefont {Goldman}\ and\ \citenamefont {Dalibard}(2014)}]{goldman2014floquet2}%
  \BibitemOpen
  \bibfield  {author} {\bibinfo {author} {\bibfnamefont {N.}~\bibnamefont {Goldman}}\ and\ \bibinfo {author} {\bibfnamefont {J.}~\bibnamefont {Dalibard}},\ }\bibfield  {title} {\bibinfo {title} {Periodically driven quantum systems: Effective hamiltonians and engineered gauge fields},\ }\href {https://doi.org/10.1103/PhysRevX.4.031027} {\bibfield  {journal} {\bibinfo  {journal} {Phys. Rev. X}\ }\textbf {\bibinfo {volume} {4}},\ \bibinfo {pages} {031027} (\bibinfo {year} {2014})}\BibitemShut {NoStop}%
\end{thebibliography}%

\newpage
\onecolumngrid
\appendix 
\clearpage
\renewcommand\thefigure{S\arabic{figure}}    
\setcounter{figure}{0} 
\renewcommand{\theequation}{S\arabic{equation}}
\setcounter{equation}{0}
\renewcommand{\thesubsection}{SM\arabic{subsection}}

\begin{center}
\textbf{\large Supplemental Material: Orbital Inverse Faraday and Cotton-Mouton Effects in Hall Fluids}
\end{center}

\section{I. High-frequency expansion}

The high-frequency regime is beyond the hydrodynamic treatment. A complementary method is to start from the microscopic Hamiltonian for the 2DEG in a constant magnetic field $B_0\hat{\bf e}_z$, coupled to the high-frequency light fields. We then consider the effective Floquet Hamiltonian as a $1/\omega$ expansion. The microscopic Hamiltonian reads
\begin{equation}
    H(t) = \frac{1}{2m} \sum_{i=1}^N \Big[\bm{p}_i + e \big( \bm{A}_0(\bm{r}_i) + \delta\bm{A}_\omega(t) \big)\Big]^2,
\end{equation}
where $m$ and $e$ are the mass and charge of carriers, $N$ is the number of particles, $\bm{p}_i$ is the momentum operator of the $i$-th particle, and $\bm{A}_0$ and $\delta\bm{A}_\omega$ are the vector potentials corresponding to the static background magnetic field and to the light, respectively,
\begin{align}
    \bm{A}_0(\bm{r}) &= \frac{B_0}{2}(-y \hat{\bf e}_x + x \hat{\bf e}_y), 
    &
    \delta\bm{A}_\omega(t) &= -\frac{1}{i\omega} \bm{E}_\omega e^{i\omega t} + \mathrm{c.c.}.
\end{align}
It is useful to introduce the kinetic momentum $\bm{\pi}_i = \bm{p}_i + e \bm{A}_0(\bm{r}_i)$ that satisfies $[\pi_i^x, \pi_j^y] =  -i\hbar e B_0 \delta_{ij}$. The Hamiltonian can then be decomposed as 
\begin{align} 
    H(t) &= \frac{1}{2m} \sum_i \big(\bm{\pi}_i + e \delta\bm{A}_\omega(t) \big)^2 
    \\
    &= H_0 + H_\omega e^{i\omega t} + H^*_{\omega} e^{-i\omega t} + H_{2\omega} e^{2i\omega t} + H^*_{2\omega} e^{-2i\omega t},
\end{align}
where
\begin{align}
    H_{0} &= \sum_{i = 1}^N \frac{\bm{\pi}_i^2}{2m} + \frac{N e^2}{m \omega^2} \abs{\bm{E}_\omega}^2, &
    H_{\omega} &= \frac{ie}{m \omega} \sum_{i=1}^N \bm{E}_\omega \cdot \bm{\pi}_i, &
    H_{2\omega} &= -\frac{N e^2}{m \omega ^2} |\bm{E}_\omega|^2.
\end{align}
The effective Floquet Hamiltonian $H_{\rm eff}$ is defined through $\exp(-iTH_{\rm eff})=\mathcal{T} \exp(-i\int_0^T \!\! \dd t \,H(t))$, where $T=2\pi/\omega$ is the period of oscillation of the light field. Its high-frequency expansion is given by \cite{rahav2003floquet1, goldman2014floquet2},
\begin{equation} \label{eqS:effective_floquet_hamiltonian}
    H_\mathrm{eff.} = H_\mathrm{L} + H_{(0)} + H_{(1)} +  H_{(2)} + \dots,
\end{equation}
where the Landau Hamiltonian is $H_L = \sum_{i = 1}^N \bm{\pi}_i^2 / (2m)$, and
\begin{align}
    H_{(1)} &= \frac{1}{\hbar \omega} [H_\omega, H^*_\omega],
    &
    H_{(2)} &= \frac{1}{2(\hbar \omega)^2} \Big( \big[H_\omega, [H_0, H^*_\omega]\big] + \text{c.c.} \Big).
\end{align}
We compute the relevant commutators as
\begin{align}
    [H_\omega, H^*_\omega] &= \frac{e^2}{m^2\omega^2} E_\omega^a E_\omega^{* b} \sum_{i, j = 1}^N [\pi_i^a, \pi_j^b] = -\frac{i\hbar Ne^3 B_0 }{m^2 \omega^2} \bm{E}_\omega \times \bm{E}_\omega^*,
    \\
    \big[H_\omega, [H_0, H^*_\omega] \big] &= \frac{e^2}{2m^3 \omega^2} E^a_\omega E^{* b}_\omega \sum_{i, j, k = 1}^N \big[\pi^a_i, [\bm{\pi}^2_j, \pi^b_k]\big] = \frac{N \hbar^2 e^4 B^2_0}{m^3 \omega^2} |\bm{E}_\omega|^2,
\end{align}
where repeated indices $a, b \in \{x,y\}$ are implicitly summed over.

The deviation from the Landau Hamiltonian to leading order $1/\omega^2$ is
\begin{equation}
    H_{(0)} = \frac{N e^2}{m\omega^2} |\bm{E}_\omega|^2,
\end{equation}
which gives a potencial quadratic in the applied field. Its gradient leads to the ponderomotive force. At the next order $1/\omega^3$ we find
\begin{equation}
    H_{(1)} = -\frac{i N e^3 B_0}{m^2 \omega^3} \bm{E}_\omega \times \bm{E}^*_\omega, 
\end{equation}
which couples linearly to the static magnetic field. Interpreting this contribution as a Zeeman term $H \sim -\bm{\mu} \cdot \bm{B}$ we find an induced out-of-plane magnetic moment, which we conveniently express as
\begin{equation}
    \mu_{(1)} = \frac{i N e^3}{m^2 \omega^3} \bm{E}_\omega \times \bm{E}^*_\omega = \frac{i A}{e \omega \rho_0} \Big[\frac{\rho_0^2 e^4}{m^2 \omega^2}\Big] \bm{E}_\omega \times \bm{E}^*_\omega,
\end{equation}
where $A$ is the area of the system and $\rho_0=N/A$ is the number density. Defining the filling fraction as $\nu=\frac{N}{B_0A/\Phi_0}$, with $\Phi_0=\frac{h}{e}$ the magnetic flux quantum, we have that $\rho_0 = e \nu B_0 / h=m \nu \omega_0/ h$, with the cyclotron frequency $\omega_0 = e B_0 / m$. Using this relation, we find that the induced magnetization density $M=\mu/A$ is given by
\begin{equation}
    M_{(1)} = \frac{i}{e \omega \rho_0} \Big[ \frac{\nu^2 e^4 \omega^2_0}{h^2 \omega^2} \Big] \bm{E}_\omega \times \bm{E}^*_\omega = \frac{i}{e \omega \rho_0} |\sigma^L_{\omega \gg \omega_0}|^2 \bm{E}_\omega \times \bm{E}^*_\omega,
\end{equation}
where we recognized the leading-order high-frequency expansion of the longitudinal conductivity $\sigma^L_\omega=\nu\frac{e^2}{h}\frac{i\omega\omega_0}{\omega^2-\omega_0^2}$.

The $1/\omega^4$ term in the effective Hamiltonian is
\begin{equation}
    H_{(2)} = \frac{N e^4 B^2_0}{m^3 \omega^4} |\bm{E}_\omega|^2 = \frac{Ne^3 B_0 \omega_0}{m^2 \omega^4} |\bm{E}_\omega|^2,
\end{equation}
which similarly corresponds to a magnetization
\begin{equation}
    M_{(2)}
    = \frac{1}{e\omega \rho_0} \Big[-\frac{\nu^2 e^4 \omega^3_0}{h^2 \omega^3} \Big] |\bm{E}_\omega|^2 
    = \frac{1}{e \omega \rho_0} {\rm Im
    }\big(\sigma_{\omega \gg \omega_0}^L \sigma_{\omega \gg \omega_0}^H\big) |\bm{E}_\omega|^2.
\end{equation}

\section{II. Nonlinear response in the QH hydrodynamics}

The hydrodynamic action of the QH fluid at filling fraction $\nu$ is given by
\begin{equation}
    S = -\int d^2x\left[\rho\left(\p_t\theta+eA_t+\frac{m{\bf v}^2}{2}\right)+V(\rho)\right],\label{eq:QHactionSup}
\end{equation}
in terms of the particle density $\rho$ and the conjugated phase variable $\theta$. The fluid velocity $\bf v$ is defined in terms of these variables as
\begin{equation}
    m {\bf v}=\nabla \theta+\frac{e}{c}{\bf A}+\frac{2\pi\hbar}{\nu}\nabla^\perp\left(\frac{\rho}{\Delta}\right),\label{eq:vHallSup}
\end{equation}
where we use the abreviated notations $v^\perp_i=\varepsilon_{ij}v^j$ and $\frac{\rho}{\Delta}(x)=\int d^2x'\frac{\log|x-x'|}{2\pi}\rho(x')$. Varying \eqref{eq:QHactionSup} over $A_\mu$ one finds that the charge density is given by
\begin{equation}
    {\bf J} = e\rho{\bf v}.\label{eq:Jvtilde}
\end{equation}
This agrees with the general constitutive relation, and justifies identifying $\bf v$ with the fluid velocity. We will make use of the fact that the vorticity of the fluid is determined by the density,
\begin{equation}
    \nabla\times{\bf v}=\frac{eB}{m}-\frac{2\pi}{m\nu}\rho,\label{eq:QHconstraint}
\end{equation}
where from this point on we use $\hbar=c=1$ units. This is known as the quantum Hall constraint. Note that we use standard vector notation for consistency with the main text, but in two dimensions it is natural to think of $\nabla\times\bf v$ as a pseudo-scalar.

The equations of motion following from the action principle \eqref{eq:QHactionSup} are the conservation laws
\begin{align}
    &\p_t\rho+\p^iJ_i=0,\label{eq:continuityQH}\\
    &\p_tJ_i+\p^jT_{ij}=\rho F_i,\label{eq:forceQH}
\end{align}
where
$T_{ij}=m\rho v_iv_j$ is the energy-momentum tensor of the QH fluid and $F_i = eE_i-ev^\perp_iB$, with $E_i = -\p_tA_i-\p_iA_t$ and $B=\nabla\times\bf A$, is the Lorentz force.

\subsection{Linear response}

We briefly review the calculation of the linear response in the QH state. We follow the hydrodynamic calculation in \cite{abanov2013effective}, while introducing a field parametrization which will be useful in the calculation of the non-linear response. In a constant external magnetic field $B_0$, the equations (\ref{eq:continuityQH},\ref{eq:forceQH}) have a homogeneous solution ${\bf v}=0$, $\rho=\rho_0$. However, one should check that ${\bf v}=0$ does correspond to a configuration of the $\theta$ variable. From the definition \eqref{eq:vHallSup}, it is enough to impose the vanishing of the rotational part of $\bf v$ which, by \eqref{eq:QHconstraint}, gives the Streda relation
\begin{equation}
    \rho_0=\nu\frac{eB_0}{2\pi}=\frac{\nu m\omega_0}{2\pi}.
\end{equation}
where $\omega_0 = \frac{eB_0}{m}$ is the cyclotron frequency. Consider the linear response to perturbing this saddle point by an additional field $\delta A_\mu$, which we decompose as
\begin{equation}
    \delta {\bf A} = \delta {\bf A}^L+\delta {\bf A}^T,
\end{equation}
where
\begin{align}
    &\delta {\bf A}^L = \nabla\left(\frac{\nabla\cdot\delta {\bf A}}{\Delta}\right), &\delta {\bf A}^T=-\nabla^\perp\left(\frac{\nabla\times\delta {\bf A}}{\Delta}\right)
\end{align}
is the Helmholz decomposition of the vector potential, plus $\delta A_t$. We will use the simplified notations
\begin{align}
    &B = \nabla\times\delta{\bf A}, &{\bf E} = -\p_t\delta{\bf A}-\nabla\delta A_t,
\end{align}
where the perturbation to the magnetic field $B$ should not be confused with the (large) background magnetic field $B_0=\nabla\times {\bf A}_0$. We find it convenient to use the Hamiltonian (temporal) gauge $\delta A_t=0$, so that
\begin{align}
    &{\bf E} = {\bf E}^L+{\bf E}^T, &{\bf E}^{L(T)}=-\p_t\delta {\bf A}^{L(T)},\\
    &B=B^T=\nabla\times \delta {\bf A}^T.
\end{align}
The usefulness of these relations comes from the following. Recall that
\begin{equation}
    {\bf J} = \nabla(\p_t P)-\nabla^\perp  M,\label{eq:jPM}
\end{equation}
where $\bf P$ and $\bf M$ are the electric polarization and magnetization densities, respectively. In terms of the coupling to probe fields,
\begin{equation}
    -\int {\bf J}\cdot \delta{\bf A} = \int\p_tP(\nabla\cdot \delta{\bf A})+\int M(\nabla\times \delta{\bf A})= \int P(\nabla\cdot {\bf E}^L)+\int MB,
\end{equation}
so that
\begin{align}
    &P = \frac{\delta S_{\rm eff}}{\delta(\nabla\cdot{\bf E}^L)}, &M = \frac{\delta S_{\rm eff}}{\delta B}.
\end{align}
Note that these formulas are particular to two dimensions, which agrees with the fact that the two functions $P$ and $M$ determine the vector field $\bf J$ by \eqref{eq:jPM}. Since we are interested in the induced orbital magnetization, we can directly vary over the external magnetic field. In the calculation of the nonlinear response, this will simplify the calculation.

Expanding the action \eqref{eq:QHactionSup} to second order around the ${\bf v}=0$, $\rho=\rho_0$ saddle point gives
\begin{align}
    S^{(2)}=\int dtd^2x\left\{\frac{1}{2}\begin{pmatrix}
        \delta\phi & \frac{\delta\rho}{\omega_0^2}
    \end{pmatrix}G^{-1}\begin{pmatrix}
        \delta\phi\\
        \frac{\delta\rho}{\omega_0^2}
    \end{pmatrix}+\frac{\nu e^2\omega_0}{4\pi}B\frac{1}{\Delta}B+v\frac{\delta\rho}{\omega_0^2}\right\},\label{eq:gaussian}
\end{align}
with
\begin{align}
    &G^{-1}=\begin{pmatrix}
        \frac{\nu\omega_0}{2\pi}\Delta & -\omega_0^2\p_t\\
        \omega_0^2\p_t & \frac{2\pi\omega_0^5}{\nu}\Delta^{-1}
    \end{pmatrix}, &v = \frac{e\omega_0^2}{\Delta}(\nabla\cdot {\bf E}^L-\omega_0B).\label{eq:Gminusv}
\end{align}
Integrating over fluctuations of the hydrodynamic fields, we find that to this order the effective response action is
\begin{align}
    &S_{\rm eff}^{(2)}[A]=\int dtd^2x\left\{-\frac{1}{2}vG_{22}v+\frac{\nu e^2\omega_0}{4\pi}B\frac{1}{\Delta}B\right\},
    &G=\frac{1}{\omega_0^3(\omega_0^2+\p_t^2)}\begin{pmatrix}
        \frac{2\pi\omega_0^4}{\nu}\Delta^{-1} & \omega_0\p_t\\
        -\omega_0\p_t & \frac{\nu}{2\pi}\Delta
    \end{pmatrix}.
\end{align}
Then we find
\begin{align}
    M =\nu\frac{e^2}{2\pi}\left[\frac{\omega_0^2}{\omega_0^2-\omega^2}\frac{\nabla\cdot {\bf E}^L}{\Delta}-\frac{\omega_0^3}{\omega_0^2-\omega^2}\frac{B}{\Delta}+\omega_0\frac{B}{\Delta}\right].
\end{align}
Similarly,
\begin{align}
    P=\nu\frac{e^2}{2\pi}\left[-\frac{\omega_0}{\omega_0^2-\omega^2}\frac{\nabla\cdot {\bf E}^L}{\Delta}+\frac{\omega_0^2}{\omega_0^2-\omega^2}\frac{B}{\Delta}\right].
\end{align}
Substituting in \eqref{eq:jPM}, and using that
\begin{align}
    &\nabla\left(\frac{\nabla\cdot {\bf E}^L}{\Delta}\right) = {\bf E}^L, &\nabla^\perp\left(\frac{B}{\Delta}\right)=\nabla^\perp\left(\frac{\nabla\times {\bf A}^T}{\Delta}\right)=-{\bf A}^T = -\frac{{\bf E}^T}{i\omega},
\end{align}
we find
\begin{align}
    {\bf J}&=\nu\frac{e^2}{2\pi}\Bigg[\frac{i\omega_0\omega}{\omega_0^2-\omega^2}{\bf E}^L-\frac{\omega_0^2}{\omega_0^2-\omega^2}\varepsilon_{kl}E^{T,l}-\frac{\omega_0^2}{\omega_0^2-\omega^2}\varepsilon_{kl}E^{L,l}+\frac{i\omega_0^3}{\omega(\omega_0^2-\omega^2)}{\bf E}^T-\frac{i\omega_0}{\omega}{\bf E}^T\Bigg]\\
    &=\nu\frac{e^2}{2\pi}\left[\frac{i\omega_0\omega}{\omega^2-\omega_0^2}{\bf E}-\frac{\omega_0^2}{\omega^2-\omega_0^2}{\bf E}^\perp\right].
\end{align}
The admittance tensor has both an imaginary longitudinal and a real Hall component. For large $\omega_0$ (topological scaling limit), the Hall response dominates.

\subsection{Second order perturbation}

Applying the notations above, the third-order terms in the action become
\begin{align}
    S^{(3)}=&-\int dtd^2x\Bigg\{\frac{e^2}{2m}\delta\rho\p_i\left(\frac{B}{\Delta}\right)\p^i\left(\frac{B}{\Delta}\right)-\frac{e}{m}\delta\rho\p^i\delta\phi\p^\perp_i\left(\frac{B}{\Delta}\right)-\frac{e}{m}\frac{2\pi}{\nu}\delta\rho\p^i\left(\frac{\delta\rho}{\Delta}\right)\p_i\left(\frac{B}{\Delta}\right)\\
    &\hspace{.6\linewidth}+\frac{\delta\rho}{2m}\left[\p_i\delta\phi+\frac{2\pi}{\nu}\p^\perp_i\left(\frac{\delta\rho}{\Delta}\right)\right]^2\Bigg\}.
\end{align}
The first line introduces corrections to the Gaussian integral \eqref{eq:gaussian},
\begin{align}
    &\delta G^{-1} = \begin{pmatrix}
        0 & -\frac{\nu e\omega_0^3}{2\pi\rho_0}\p^\perp_i\left(\frac{B}{\Delta}\right)\p^i\\
        \frac{\nu e\omega_0^3}{2\pi\rho_0}\p^\perp_i\left(\frac{B}{\Delta}\right)\p^i & \frac{e\omega_0^5}{\rho_0}\p_i\left(\frac{B}{\Delta}\right)\frac{\p^i}{\Delta}
    \end{pmatrix},
    &\delta v = \frac{\nu e^2\omega_0^3}{4\pi\rho_0}\p_i\left(\frac{B}{\Delta}\right)\p^i\left(\frac{B}{\Delta}\right),
\end{align}
while the second line introduces interactions, which we will consider later. We are only interested in the generation of magnetization by the transverse part of the electric field,
\begin{equation}
    M = \frac{\delta S_{\rm eff}}{\delta B}\Bigg|_{E^L=0},
\end{equation}
which we keep in mind when considering the relevant terms in
\begin{align}
    S_{\rm eff}^{(3)}[A]=\int dtd^2x\left\{-\frac{1}{2}(v+\delta v)(G^{-1}+\delta G^{-1})^{-1}_{22}(v+\delta v)+\frac{1}{2}vG_{22}v\right\}
    =\int dtd^2x\left\{\frac{1}{2}v(G^2\delta G^{-1})_{22}v-vG_{22}\delta v\right\}.\label{eq:cubicEB}
\end{align}
We therefore consider only the terms cubic in $B$. Using the short-hand notation
\begin{equation}
    G^2 = \frac{1}{\omega_0^6(\omega_0^2+\p_t^2)^2}\begin{pmatrix}
        \frac{4\pi^2\omega_0^8}{\nu^2}\Delta^{-2}-\omega_0^2\p_t^2 & \left(\frac{\nu}{2\pi}\Delta+\frac{2\pi\omega_0^4}{\nu}\Delta^{-1}\right)\omega_0\p_t\\
        -\left(\frac{\nu}{2\pi}\Delta+\frac{2\pi\omega_0^4}{\nu}\Delta^{-1}\right)\omega_0\p_t & \frac{\nu^2}{4\pi^2}\Delta^2-\omega_0^2\p_t^2
    \end{pmatrix},
\end{equation}
we find, from the $G^2\delta G^{-1}$ term,
\begin{align}
    &\frac{\nu^2e^3\omega_0^4}{8\pi^2\rho_0}\int dtd^2x\Bigg\{\left(\frac{B}{\Delta}\right)\frac{1}{(\omega_0^2+\p_t^2)^2}\Bigg[\left(\Delta+\frac{4\pi^2\omega_0^4}{\nu^2}\Delta^{-1}\right)\p_t\p^\perp_i\left(\frac{B}{\Delta}\right)\p^i\\
    &\hspace{.5\linewidth}+\omega_0\left(\Delta^2-\omega_0^2\p_t^2\right)\p_i\left(\frac{B}{\Delta}\right)\frac{\p^i}{\Delta}\Bigg]\left(\frac{B}{\Delta}\right)\Bigg\}.\label{eq:Gcorrectionterms}
\end{align}
To compute the DC magnetization, we note that
\begin{equation}
    \frac{\delta}{\delta \alpha(\omega,k)}\int dtd^2x\alpha(t,x)\beta(t,x)\gamma(t,x)=\int\frac{d\omega'}{2\pi}\frac{d^2k'}{(2\pi)^2}\beta(\omega',k')\gamma(-\omega-\omega',-k-k'),
\end{equation}
where we use the same name for the functions $\alpha,\beta,\gamma$ and their Fourier transforms. We are interested in the DC magnetization $M(0,k)$, which allows us to drop a few terms in \eqref{eq:Gcorrectionterms}. From the first line, the surviving term comes from varying over the first and last factors of $B$,
\begin{align}
    M_A(0,k)=-\frac{1}{k^2}\int\frac{d\omega}{2\pi}\frac{d^2q}{(2\pi)^2}\frac{\sigma_H^2(\omega)}{ie\rho_0\omega}\left(q^2+\frac{4\pi^2\omega_0^4}{\nu^2q^2}\right)E^T(\omega,q)\times E^{T}(-\omega,-k-q).
\end{align}
From the second line there are two terms. Varying over the first and last factors of $B$ gives
\begin{align}
    M_B(0,k)=-\frac{1}{k^2}\int\frac{d\omega}{2\pi}\frac{d^2q}{(2\pi)^2}\frac{\omega_0^2{\rm Im}[\sigma_L^*\sigma_H](\omega)}{e\rho_0\omega^3}\frac{q^4+\omega_0^2\omega^2}{(k+q)^2}E^T(\omega,q)\cdot E^{T}(-\omega,-k-q),
\end{align}
while varying over the second factor of $B$ gives
\begin{align}
    M_C(0,k)&=\frac{\nu^2e^3\omega_0^4}{8\pi^2\rho_0}\int\frac{d\omega}{2\pi}\frac{d^2q}{(2\pi)^2}\frac{-q^\perp\cdot E^T(\omega,q)}{\omega q^2}\frac{k^4}{\omega_0^3}\frac{ik_i}{-k^2}\left[-\varepsilon^{ij}\frac{E^T_j(-\omega,-k-q)}{i\omega(k+q)^2}\right].
\end{align}
This contribution depends explicitly on $\p^\perp\cdot E^T\propto\p_tB$, which at this point we set to zero.

Let us consider now the $\delta v$ term in \eqref{eq:cubicEB},
\begin{equation}
    \frac{\nu^2e^3\omega_0^3}{8\pi^2\rho_0}\int dtd^2x\left(\frac{B}{\Delta}\right)\frac{1}{\omega_0^2+\p_t^2}\Delta\left[\p^\perp_i\left(\frac{B}{\Delta}\right)\p^{\perp i}\left(\frac{B}{\Delta}\right)\right].
\end{equation}
Varying over the first $B$ gives
\begin{align}
    M_D(0,k)=\int\frac{d\omega}{2\pi}\frac{d^2q}{(2\pi)^2}\frac{(\omega_0^2-\omega^2){\rm Im}[\sigma_L^*\sigma_H](\omega)}{2e\rho_0\omega^3}E^T(\omega,q)\cdot E^{T}(-\omega,-k-q),
\end{align}
and over the second/third factor gives
\begin{align}
    M_E(0,k)&=\frac{\nu^2e^3\omega_0^3}{4\pi^2\rho_0}\int\frac{d\omega}{2\pi}\frac{d^2q}{(2\pi)^2}\frac{-q^\perp\cdot E^T(\omega,q)}{\omega q^2}\frac{-q^2}{\omega_0^2-\omega^2}\frac{k^\perp_i}{-k^2}\frac{E^{T,i}(-\omega,-k-q)}{i\omega},
\end{align}
which we disregard for the same reasons as $M_C$.

We consider now the effect of interactions. To leading order, these give
\begin{equation}
    \Big\langle\int dtd^2x\frac{\delta\rho}{2m}\left[\p_i\delta\phi+\frac{2\pi}{\nu}\p^\perp_i\left(\frac{\delta\rho}{\Delta}\right)\right]^2\Big\rangle.
\end{equation}
The first term gives
\begin{align}
    -\frac{\nu\omega_0^3}{4\pi\rho_0}\int dtd^2x(G_{22}*v)(x)(G_{12}*\p^iv)(x)(G_{12}*\p_iv)(x),\label{eq:firstint}
\end{align}
where we dropped the disconnected terms, which are canceled by normalization. The transverse-field contributions are then
\begin{align}
    \frac{\nu e^3\omega_0^3}{4\pi\rho_0}\int dtd^2x\left(\frac{1}{\omega_0^2+\p_t^2}\frac{\nu}{2\pi}B\right)\left(\frac{1}{\omega_0^2+\p_t^2}\omega_0E^{T,i}\right)\left(\frac{1}{\omega_0^2+\p_t^2}\omega_0E^T_i\right),
\end{align}
which leads to
\begin{align}
    M_F(0,k)=-\int\frac{d\omega}{2\pi}\frac{d^2q}{(2\pi)^2}\frac{{\rm Im}[\sigma_L^*\sigma_H](\omega)}{2e\rho_0\omega}E^T(\omega,q)\cdot E^{T}(-\omega,-k-q).
\end{align}

Repeating the same steps, we find
\begin{align}
    M_G(0,k)=\int\frac{d\omega}{2\pi}\frac{d^2q}{(2\pi)^2}\frac{i\sigma_H^2(\omega)}{2e\rho_0\omega}E^T(\omega,q)\times E^{T}(-\omega,-k-q),
\end{align}
and
\begin{align}
    M_H(0,k)=-\int\frac{d\omega}{2\pi}\frac{d^2q}{(2\pi)^2}\frac{\omega_0^2{\rm Im}[\sigma_L^*\sigma_H](\omega)}{2e\rho_0\omega^3}E^T(\omega,q)\cdot E^{T}(-\omega,-k-q).
\end{align}

Let us finally organize these contributions. There are two non-local contributions,
\begin{align}
    M_A(0,k)+M_B(0,k)&=-\frac{1}{k^2}\int\frac{d\omega}{2\pi}\frac{d^2q}{(2\pi)^2}\Bigg[\frac{\sigma_H^2(\omega)}{ie\rho_0\omega}\left(q^2+\frac{4\pi^2\omega_0^4}{\nu^2q^2}\right)E^T(\omega,q)\times E^{T}(-\omega,-k-q)\\
    &\hspace{.25\linewidth}+\frac{\omega_0^2{\rm Im}[\sigma_L^*\sigma_H](\omega)}{e\rho_0\omega^3}\frac{q^4+\omega_0^2\omega^2}{(k+q)^2}E^T(\omega,q)\cdot E^{T}(-\omega,-k-q)\Bigg].
\end{align}
Note that these can be written as $\Delta (M_A+M_B)=...$, whose solution depends on boundary conditions. Then there are two contributions which vanish for zero extra additional magnetic flux through the Hall plane $B=0$,
\begin{equation}
    (M_C(0,k)+M_E(0,k))|_{B=0}=0.
\end{equation}
Finally, the local contributions to the magnetization give
\begin{align}
    M_D+M_F+M_H&=\int\frac{d\omega}{2\pi}\frac{d^2q}{(2\pi)^2}\left[\frac{(\omega_0^2-\omega^2)}{2e\rho_0\omega^3}-\frac{1}{2e\rho_0\omega}-\frac{\omega_0^2}{2e\rho_0\omega^3}\right]{\rm Im}[\sigma_L^*\sigma_H](\omega)E^T(\omega,q)\cdot E^{T}(-\omega,-k-q)\\
    &=\int\frac{d\omega}{2\pi}\frac{d^2q}{(2\pi)^2}\frac{{\rm Im}[\sigma_L^*\sigma_H](\omega)}{e\rho_0\omega}E^T(\omega,q)\cdot E^{T}(-\omega,-k-q)
\end{align}
and
\begin{align}
    M_G(0,k)=\int\frac{d\omega}{2\pi}\frac{d^2q}{(2\pi)^2}\frac{i\sigma_H^2(\omega)}{2e\rho_0\omega}E^T(\omega,q)\times E^{T}(-\omega,-k-q).
\end{align}

\section{III. Density profiles}

Reverting back to the action \eqref{eq:QHactionSup}, we see that the density is given by
\begin{align}
    &\langle\rho\rangle = -\frac{1}{e} \frac{\delta}{\delta A_t}S_{\rm eff}[A], &S_{\rm eff}[A]=-\int dt d^2x\rho_0(\p_t\theta_0+eA_t)+S_{\rm eff}^{(2)}[A]+S_{\rm eff}^{(3)}[A]+...,
\end{align}
where $\theta_0$ is determined by setting $\vec v=0$ in \eqref{eq:vHallSup} and $S_{\rm eff}^{(1)}$ vanishes by the equations of motion. Notice also that the variation of $S_{\rm eff}^{(2)}$ gives contributions linear in the fields, which are therefore oscillating. Thus the DC part of $\rho(\omega,k)$ is
\begin{equation}
    \rho(0,k)=\rho_0-\frac{1}{e}\frac{\delta}{\delta A_t}S_{\rm eff}^{(3)}[A](0,k)+...,
\end{equation}
and $A_t$ appears in the source term \eqref{eq:Gminusv},
\begin{equation}
    v = \frac{e\omega_0^2}{\Delta}(\p\cdot E-\omega_0B)=-e\omega_0^2A_t+...
\end{equation}
Let us evaluate the relevant contributions in an analogous way to the calculation of $M(0,k)$. From the $G^2\delta G^{-1}$ term in \eqref{eq:cubicEB},
\begin{align}
    &\frac{\nu^2e^3\omega_0^3}{8\pi^2\rho_0}\int dtd^2x\Bigg\{A_t\frac{1}{(\omega_0^2+\p_t^2)^2}\Bigg[\left(\Delta+\frac{4\pi^2\omega_0^4}{\nu^2}\Delta^{-1}\right)\p_t\p^\perp_i\left(\frac{B}{\Delta}\right)\p^i\\
    &\hspace{.5\linewidth}+\omega_0\left(\Delta^2-\omega_0^2\p_t^2\right)\p_i\left(\frac{B}{\Delta}\right)\frac{\p^i}{\Delta}\Bigg]\left(\frac{B}{\Delta}\right)\Bigg\},
\end{align}
and a similar term with $A_t$ as the last factor. From the first line, we get the contributions
\begin{align}
    \rho_A(0,k)&=-\frac{\nu^2e^2\omega_0^3}{4\pi^2\rho_0}\int\frac{d\omega}{2\pi}\frac{d^2q}{(2\pi)^2}\frac{1}{(\omega_0^2-\omega^2)^2}\left(-q^2-\frac{4\pi^2\omega_0^4}{\nu^2q^2}\right)[-E^T_i(\omega,q)]\left[\varepsilon^{ij}\frac{E^T_j(-\omega,-k-q)}{i\omega}\right]\\
    &=-\frac{1}{e\omega_0}\int\frac{d\omega}{2\pi}\frac{d^2q}{(2\pi)^2}\frac{\sigma_H^2(\omega)}{ie\rho_0\omega}\left(q^2+\frac{4\pi^2\omega_0^4}{\nu^2q^2}\right)E^T(\omega,q)\times E^{T}(-\omega,-k-q)\\
    &=\frac{k^2}{e\omega_0}M_A(0,k).
\end{align}

From the second line, we get
\begin{align}
    \rho_B(0,k)&=-\frac{\nu^2e^2\omega_0^3}{4\pi^2\rho_0}\int\frac{d\omega}{2\pi}\frac{d^2q}{(2\pi)^2}\frac{1}{(\omega_0^2-\omega^2)^2}\omega_0\left(q^4+
    \omega_0^2\omega^2\right)\frac{E^T_i(\omega,q)}{i\omega}\frac{E^{T,i}(-\omega,-k-q)}{i\omega(k+q)^2}\\
    &=-\frac{1}{e\omega_0}\int\frac{d\omega}{2\pi}\frac{d^2q}{(2\pi)^2}\frac{\omega_0^2{\rm Im}[\sigma_L^*\sigma_H](\omega)}{e\rho_0\omega^3}\frac{q^4+\omega_0^2\omega^2}{(k+q)^2}E^T(\omega,q)\cdot E^{T}(-\omega,-k-q)\\
    &=\frac{k^2}{e\omega_0}M_B(0,k).
\end{align}
Note that there is no correspondent of the contribution $M_C(0,k)$, but this was in fact zero for vanishing perturbation to the magnetic flux.

Next we consider the terms coming from $\delta v$ in \eqref{eq:cubicEB},
\begin{equation}
    \frac{\nu^2e^3\omega_0^2}{8\pi^2\rho_0}\int dtd^2x A_t\frac{1}{\omega_0^2+\p_t^2}\Delta\left[\p^\perp_i\left(\frac{B}{\Delta}\right)\p^{\perp i}\left(\frac{B}{\Delta}\right)\right],
\end{equation}
which gives, by the same steps,
\begin{equation}
    \rho_D(0,k)=\frac{k^2}{e\omega_0}M_D(0,k),
\end{equation}
and similarly there is no $\rho_E$ contribution, but $M_E$ vanishes for zero flux.

Finally, we check the interaction terms. The first term gives
\begin{align}
    \frac{\nu\omega_0}{4\pi\rho_0}\langle(\delta\rho\p_i\delta\phi\p^i\delta\phi)(x)\rangle
    =-\frac{\nu\omega_0^3}{4\pi\rho_0}\int dtd^2x(G_{22}*v)(x)(G_{12}*\p^iv)(x)(G_{12}*\p_iv)(x)
\end{align}
The transverse-field but zero flux contribution comes from
\begin{align}
    \frac{\nu e^3\omega_0^2}{4\pi\rho_0}\int dtd^2x\left(\frac{1}{\omega_0^2+\p_t^2}\frac{\nu}{2\pi}\Delta A_t\right)\left(\frac{1}{\omega_0^2+\p_t^2}\omega_0E^{T,i}\right)\left(\frac{1}{\omega_0^2+\p_t^2}\omega_0E^T_i\right),
\end{align}
which gives a contribution
\begin{equation}
    \rho_F(0,k) = \frac{k^2}{e\omega_0}M_F(0,k),
\end{equation}
and in the same way it is easy to check that the remaining contributions are
\begin{align}
    &\rho_G(0,k) = \frac{k^2}{e\omega_0}M_G(0,k), &\rho_H(0,k) = \frac{k^2}{e\omega_0}M_H(0,k).
\end{align}

It follows that all non-vanishing DC contributions satisfy
\begin{equation}
    \rho(0,k) = \frac{k^2}{e\omega_0}M(0,k).
\end{equation}
Recalling that our calculations refer to $\rho-\rho_0$, we find that the total DC density is given by the modified Streda relation
\begin{equation}
    \rho = \frac{\nu e}{2\pi}B_0 - \frac{1}{e\omega_0}\Delta M.
\end{equation}

\end{document}